\documentclass[12pt,notitlepage,a4paper]{article}

\pdfoutput=1

\usepackage{a4wide}
\usepackage{epsfig}
\usepackage{color,graphicx}
\usepackage{cite}
\usepackage{amssymb,amsmath}
\usepackage{jheppub}

\newcommand{\be}{\begin{equation}}
\newcommand{\ee}{\end{equation}}
\newcommand{\bea}{\begin{eqnarray}}
\newcommand{\eea}{\end{eqnarray}}

\begin{document}

\begin{center}  

\vskip 2cm 

\centerline{\Large {\bf\boldmath Generalized symmetries and the dimensional reduction}}
\centerline{\Large {\bf\boldmath of 6d $so$ SCFTs}}

\vskip 1cm

\renewcommand{\thefootnote}{\fnsymbol{footnote}}

   \centerline{
    {\large \bf Gabi Zafrir${}^{a,b}$}\footnote{gabi.zafrir@oranim.ac.il}}
      
\vspace{1cm}
\centerline{{\it ${}^a$ Department of Physics, University of Haifa at Oranim, Kiryat Tivon 36006, Israel}}
\centerline{{\it ${}^b$ Haifa Research Center for Theoretical Physics and Astrophysics, University of Haifa,}}
\centerline{{\it Haifa 3498838, Israel}}
\vspace{1cm}

\end{center}

\vskip 0.3 cm

\setcounter{footnote}{0}
\renewcommand{\thefootnote}{\arabic{footnote}}   
   
\begin{abstract}

We consider the dimensional reduction on a torus of the family of 6d $(1,0)$ SCFTs UV completing an $so(N)$ gauge theory with $N-8$ vector hypermultiplets. These SCFTs are known to possess a rich structure of discrete symmetries, notably 0-form and 1-form symmetries, which often merge to form a higher group structure, both split and non-split. We investigate what happens to this symmetry structure once the theory is reduced on a circle to 5d and on a torus to 4d, especially when a non-trivial Stiefel-Whitney class for the flavor symmetry is turned on. Unlike in Lagrangian theories, here the 1-form symmetries of the 6d theory reduce to non-trivially acting 1-form and 0-form symmetries, and the original higher group structure leads to an extension of the 0-form symmetries.     

\end{abstract}
 
 \newpage
 
\tableofcontents

\section{Introduction}
\label{sec:intro}

Symmetries play a fundamental role in physics. An incomplete list of their benefits includes: providing conservation laws in classical physics, selection rules in quantum physics and constraining the IR behavior in quantum field theory (QFT) through their 't Hooft anomalies. Recently we have seen a renewed interest in the subject of symmetries, motivated by the discovery of certain novel symmetry structures, now collectively referred to as generalized symmetries. These notably include symmetries acting on extended operators, now called higher form symmetries, as well as symmetries which do not form a group, now called non-invertible symmetries (see \cite{Schafer-Nameki:2023jdn,Bhardwaj:2023kri,Shao:2023gho} for some recent reviews on generalized symmetries).

More specifically, in contrast to standard symmetries,  higher form symmetries do not act on local operators but can act on extended operators, where a $p$-form symmetry naturally acts on $p$-dimensional objects. Furthermore, for the higher form symmetry to act consistently, it is important that the extended operator cannot end on lower dimensional operators. As such, higher form symmetries describe symmetries acting on probe particles or defects in a quantum field theory system. The presence of such symmetries can be used to constrain the IR dynamics of a physical system, notably through their 't Hooft anomalies (see for instance \cite{Gaiotto:2017yup,Cordova:2019bsd}), and as such are of particular interest.     

Additionally, higher form symmetries of differing degrees can mix and form a larger structure, called higher group structure \cite{Cordova:2018cvg,Benini:2018reh}. This generalizes the notion of a group extension, where two symmetries of the same form degree mix to form a larger group. The simplest generalization, called a 2-group, involves an extension of a 0-form symmetry by a 1-form symmetry. It is convenient to describe such a structure in terms of background fields. Specifically, if we introduce background fields, $B$ for the 1-form symmetry and $A$ for the 0-form symmetry, then the 2-group can be expressed by a relation of the form\footnote{This is analogous to what happens in a standard non-split group extension. For example, consider the group $\mathbb{Z}_4$ thought of as a central extension of $\mathbb{Z}_2$ by $\mathbb{Z}_2$. We can associate with it a background field $X$, which should be closed (mod 4) as it is a discrete symmetry. We can formally express it in terms of the background fields for the two $\mathbb{Z}_2$ symmetries as: $X = A_Q + 2 A_H$, with $A_Q$ and $A_H$ the background fields for the quotient and subgroup $\mathbb{Z}_2$'s, respectively. The closeness of $X$ now enforces that: $\delta A_H = \frac{1}{2}\delta A_Q = \beta A_Q$, where $\beta$ is the Bockstein homomorphism associated with the extension. As such, due to the extension, the background field associated with the subgroup is not closed, rather its "field strength" is identified with a 2-form made from the background field associated with the quotient. This works similarly in a 2-group but now the subgroup is a 1-form symmetry instead.}:

\be
\delta B = f(A) 
\ee
where $f(A)$ is some 3-form made of $A$, and we have assumed that both the 0-form and 1-form symmetries are discrete\footnote{There are similar relations also for 2-groups with continuous symmetries, but now involving the field strength rather than the connection which is usually not closed, see for instance \cite{Cordova:2018cvg}.}. Similar relations hold in the general higher group case with $B$ replaced by a background connection for some higher form symmetry and $f(A)$ replaced by some function of background connections of symmetries with lesser form degree. Such higher group symmetries have enjoyed considerable attention recently, mostly concerned with their appearance and application in QFT systems in various dimensions, see \cite{Cordova:2020tij,Hsin:2020nts,Lee:2021crt,Apruzzi:2021vcu,Apruzzi:2021mlh,Mekareeya:2022spm,Cordova:2022qtz,DelZotto:2022joo,Cvetic:2022imb,Hubner:2022kxr,Bhardwaj:2023zix,Cvetic:2024dzu} for an incomplete list of references.
 
The resulting structure then provides additional information regarding the symmetries of the underlying theory and can have interesting physical implications. As an example, consider a standard group extension, where a group $G$ is constructed from two other groups, $H$ and $Q$, such that $H$ is a normal subgroup and $Q$ is the quotient under it. If $G$ is not a direct product, then $Q$ is not a normal subgroup, and likewise $H$ is not a quotient group. Recall that in an RG flow a certain normal subgroup of the global symmetry may act trivially in the IR, and the non-trivially acting symmetry is reduced to the quotient group. However, a quotient cannot act trivially unless it is also a normal subgroup. In that sense, a group extension provides restrictions on which symmetries can act trivially in the IR, not dissimilar from a 't Hooft anomaly. Higher group structure provides similar restrictions, and are thus interesting to study.

A natural question underlying the study of generalized symmetries is what constraints can these impose on RG flows. Here we will be particularly concerned with RG flows across dimensions, more specifically, dimensional reduction. That is, we will be interested in the case where we have a $D$-dimensional quantum field theory system that is reduced on a $d$-dimensional compact manifold to yield a new $D-d$ dimensional QFT system. Dimensional reduction provides an important tool to realize and study QFT systems, especially at strong coupling. The classic example being the class S construction \cite{Gaiotto:2009we,Gaiotto:2009hg}, where 4d $\mathcal{N}=2$ SCFTs are realized through the dimensional reduction of a 6d $(2,0)$ theory on a Riemann surface. This construction can be used to realize and study many so-called non-Lagrangian 4d SCFTs that are difficult to tackle by other methods. By now, similar constructions also exist for the dimensional reduction on Riemann surfaces of 6d $(1,0)$ SCFTs \cite{Gaiotto:2015usa,Razamat:2016dpl,Bah:2017gph,Kim:2017toz,Kim:2018bpg,Kim:2018lfo,Razamat:2018gro,Zafrir:2018hkr,Razamat:2019mdt,Chen:2019njf,Pasquetti:2019hxf,Razamat:2019ukg,Razamat:2020bix,Sabag:2020elc,Hwang:2021xyw,Sabag:2022hyw,Kim:2023qbx}, 5d SCFTs \cite{Sacchi:2021afk,Sacchi:2021wvg,Sacchi:2023rtp,Sacchi:2023omn} and 4d $\mathcal{N}=1$ SCFTs \cite{Kutasov:2013ffl,Kutasov:2014hha,Putrov:2015jpa,Gadde:2015wta,Sacchi:2020pet,Nawata:2023aoq,Jiang:2024ifv,Amariti:2024usp}, at least in certain cases (see cited papers for an incomplete list of references). 

An interesting question then is what is the relation between the symmetries of the higher and lower dimensional QFT systems. For standard symmetries, we generally expect for the lower dimensional theory to inherit the symmetries of the higher dimensional one\footnote{There are two notable caveats here. First, it is possible for symmetries not present in the UV to emerge in the IR. These are usually refered to as accidental symmetries. Second, it is possible for symmetries present in the UV to act trivially in the IR.}. Furthermore, it is known that the 't Hooft anomalies of the symmetries are also related by integrating the anomaly polynomial (for continuous symmetries) \cite{Benini:2009mz,Alday:2009qq} or anomaly theory (for discrete symmetries) \cite{Sacchi:2023omn} of the higher dimensional theory on the compact surface. This provides a lot of information about the lower dimensional theory extracted from its higher dimensional parent, which can be used to infer various properties of the lower dimensional theory. In certain cases and under certain assumptions, such information can be used to conjecture a Lagrangian description for the resulting lower dimensional theory, see for instance \cite{Razamat:2019vfd,Razamat:2020bix}.  

Similar relations are expected to hold also for higher form symmetries, though now the situation is richer, as a higher form symmetry reduces to a symmetry of the same form degree, but also to one of a lower degree. For instance, consider a circle reduction of a QFT system with a 1-form symmetry acting on certain line operators in the theory. When reduced on a circle, the line operators of the higher dimensional theory reduce to both line operators and local operators (from line operators wrapping the circle) of the lower dimensional theory. As such, the original 1-form symmetry reduces to both a 1-form symmetry of the lower dimension theory, acting on the reduced line operators, and to a 0-form symmetry acting on the line operators wrapping the circle. As such, in the presence of higher form symmetries, the symmetry structure becomes more involved upon dimensional reduction. This is further complicated in the presence of higher structure, which should then reduce to similar structure involving the reduced groups. It is interesting to better understanding the relation between the symmetries and their implications, especially in the presence of a non-trivial interplay between zero and higher form symmetries.  

However, in many cases the generically expected rich structure seems to trivialize as some of the expected symmetries do not manifest in the lower dimensional theory, for instance, ending up acting trivially. For example, this usually occurs in Lagrangian theories where the electric 1-form symmetry reduces to an electric 1-form symmetry while the magnetic $d$-form symmetry reduces to a $d-1$ form symmetry, with the other expected symmetries not surviving the reduction\footnote{Specifically, the expected electric 0-form symmetry appears to be broken spontaneously, while the expected magnetic $d$-form symmetry appears to act trivially \cite{Nardoni:2024sos}}. This raises the question of whether interesting examples exhibiting the expected rich structure can be found.

The purpose of this article is to explore a specific example exhibiting such a behavior. This can then be used to explore how generalized symmetry structures, notably higher group structure, reduce in such cases, which in turn allow us to better understand the relation between symmetries under dimensional reduction. More specifically, in dimensional reduction one can also turn on various holonomies or fluxes on the compact surfaces, and their presence may change the fate of the considered symmetry structure under the reduction. For instance, in \cite{Nardoni:2024sos} it was argued that reducing a 2-group, involving continuous 0 and 1-form symmetries, on a 2-sphere in the presence of flux for the flavor symmetry on the 2-sphere leads to the breaking of said flavor symmetry to a discrete group in lower dimensions. Part of the motivation of this work is to search for a similar phenomena involving a 2-group with discrete symmetries. 

Specifically, the example we shall consider involves a family of 6d $(1,0)$ SCFTs, which are strongly coupled non-Lagrangian theories and as such can evade the usual behavior of Lagrangian models. While the SCFTs themselves have no Lagrangian description, as common in 6d $(1,0)$ SCFTs, one can go on the tensor branch where the theory has an effective description as a 6d gauge theory. For the specific model considered here, the 6d gauge theory in question consists of an $so(N)$ vector multiplet and $N-8$ hypermultiplets in the vector representation. This family of 6d SCFTs is known to possess a rich global symmetry structure, consisting of 0-form symmetries, both discrete and continuous, discrete 1-form symmetries and a discrete self-dual 2-form symmetry, which often mix together in a higher group structure (both split and non-split).   

Following previous works, notably \cite{Hayashi:2015vhy}, we can employ the known brane construction of this family of 6d SCFTs, given in \cite{Hanany:1997gh,Brunner:1997gk}, to understand its dimensional reduction on circles to 5d and 4d. These can be further generalized by the inclusion of fluxes, notably of Stiefel-Whitney type. More specifically, the 6d SCFT has a $usp(2N-16)$ flavor symmetry algebra, which for $N=2n$ is thought to globally be $PUSp(4n-16) = USp(4n-16)/\mathbb{Z}_2$. We can then consider the torus reduction in the presence of a non-trivial Stiefel-Whitney class. This is especially interesting as the Stiefel-Whitney class of the $PUSp(4n-16)$ symmetry, $w_2(PUSp(4n-16))$, is involved in a 2-group structure of the 6d SCFT, and we can inquire what is the fate of the 2-group structure when the dimensional reduction is done in the presence of a non-trivial Stiefel-Whitney class. 

We also note that the $N=8$ case is special as there is no continuous flavor symmetry in this case. However, the 6d SCFT in this case enjoys a large discrete symmetry group, consisting of a $\mathbb{Z}_2 \times \mathbb{Z}_2$ 1-form symmetry and an $S_3$ 0-form triality symmetry with a non-trivial interplay between the two (forming a split 2-group). Finally, the 6d SCFTs in this family possess a self-dual 2-form symmetry that mixes with its 1-form symmetry through a 3-group structure. We hope we managed to quickly convey the richness of the symmetry structure enjoyed by these 6d SCFTs, and below we explore what happens to this structure upon dimensional reduction on a circle to 5d and on a torus to 4d.

One interesting observation we find in the dimensional reduction is that the 1-form symmetries reduce also to 0-form symmetries, and that the interplay between the 1-form and 0-form symmetries in the 6d theory reduces to a similar one between the 0-forms symmetries of the 5d and 4d theories. For example, in the case of the $so(8)$ 6d SCFT, the split 2-group reduces to a split extension, that is a semi-direct product, upon dimensional reduction. As such, while the 6d SCFT has an $S_3$ 0-form symmetry, its reduction on a circle has an $S_4 = (\mathbb{Z}_2 \times \mathbb{Z}_2) \rtimes S_3$ global symmetry, due to the above extension, and similarly its torus reduction acquires an $\mathbb{Z}^4_2 \rtimes S_3$ global symmetry. 

Similar phenomena also occurs for a non-split extension. For instance, we find that the 6d 2-form symmetry wrapping the torus gives a 0-form symmetry in 4d and that its 3-group structure reduces to a non-split extension between this 0-form symmetry and the 0-form symmetries coming from the 6d 1-form symmetries wrapping various cycles of the torus. In the case of the $so(8)$ 6d SCFT, this leads to the previously mentioned $\mathbb{Z}^4_2 \rtimes S_3$ global symmetry being further extended to $W(D_4)$, the Weyl group of $so(8)$. Similarly, we find that if we reduce the $so(2n)$ SCFT (for $n>4$) on a circle with a certain flavor twist, such that $\int_{S^1} w_2(PUSp(4N-16)) \neq 0$, the non-split extension involving $w_2$ reduces to a non-split extension of the 0-form symmetries. Specifically, in the case considered, a $\mathbb{Z}_2 \times \mathbb{Z}_2$ 0-form symmetry, coming form the 0-form symmetry of the 6d SCFT, combines with a $\mathbb{Z}_2$ symmetry coming from the 1-form symmetry of the 6d SCFT to form a dihedral symmetry, $D_4$, in the 5d theory. We also observe that when reducing on the torus with $\int_{S^1} w_2(PUSp(4N-16)) = 1$, the 0-form symmetries participating in the 2-group appear to be broken in 4d. This generalizes results observed in \cite{Nardoni:2024sos} for continuous 2-groups to the discrete case. Overall, we see that we can learn a lot about the discrete symmetry structure of the resulting 5d ad 4d theories, from the symmetry structure of their parent 6d SCFT. We also see that much of the symmetry structure of the 0-form symmetries of the 5d and 4d theories is encoded in the higher form symmetries, and associated higher structure, of their parent 6d SCFT.

The structure of of the paper is as follows. We begin in section \ref{sec:so8} with the $N=8$ case. As mentioned, this case exhibits several unique properties, and as such warrant a separate discussion.  We then move on in section \ref{sec:soN} to discuss the reduction for generic $N$. Here we first consider the standard circle reduction, after which we move to the main part: the torus reduction with a non-trivial Stiefel-Whitney class. We finally end in section \ref{sec:concl} with some conclusions. The text is supplemented by a few appendices. Specifically, appendix \ref{App:5dSU4} discusses the matching of the 1-form symmetries in certain 5d dualities. In addition to their general interest, these provide a simpler example of the same matching between the 1-form symmetry of the 6d SCFT reduced on a circle and the resulting 5d theory, that appears profusely in this paper. Appendices \ref{App:D4} and \ref{App:WD4} discuss some aspects of discrete groups that are of use in this paper, specifically, the properties of the dihedral group $D_4$ and the Weyl group of $so(8)$, respectively.    

\section{Torus reduction of the 6d $so(8)$ SCFT}
\label{sec:so8}

We shall begin our study of the reduction of the 6d $so(N)$ SCFTs with the case of $N=8$. This case possesses a remarkably rich set of discrete symmetries compared to the higher $N$ cases and so shall be treated separately. We begin with a short review of the properties of this 6d SCFT:

\begin{itemize}
 \item As mentioned, the 6d SCFT has a 1d tensor branch on a generic point of which the low-energy description is given by a pure $so(8)$ gauge theory. Here we shall make a specific choice of global structure such that the gauge group is $Spin(8)$. This fixes some of the discrete symmetry spectrum of the SCFT. Other choices can then be reached by gauging discrete symmetries. The SCFT also has a 1d Higgs branch whose geometry is $\mathbb{C}^4/\hat{D}_{4}$.  
 \item It has no continuous global symmetries save for the 6d conformal symmetry. The anomalies under this symmetry can be encoded in an anomaly polynomial, see \cite{Razamat:2018gro} for instance for the exact expression.
 \item The SCFT should possess discrete global symmetries. Specifically, it should have an $S_3$ symmetry associated with the outer automorphism of $Spin(8)$.
 \item Additionally it has discrete 1-form symmetries, associated with the $\mathbb{Z}_2 \times \mathbb{Z}_2$ center of $Spin(8)$. These act on the vector and spinor Wilson lines, and are thought to originate from 1-form symmetries of the 6d SCFT (see for instance \cite{Apruzzi:2020zot,Apruzzi:2022dlm}). This leads us to expect that the 6d SCFT has a $\mathbb{Z}_2 \times \mathbb{Z}_2$ 1-form symmetry. Note that the action in the gauge theory also implies a non-trivial action of the $S_3$ symmetry on the $\mathbb{Z}_2 \times \mathbb{Z}_2$ 1-form symmetry, where the action is by the permutation of the three $\mathbb{Z}_2$ subgroups of $\mathbb{Z}_2 \times \mathbb{Z}_2$ (this structure is sometimes refereed to as a split 2-group).   
 \item Finally, the SCFT should also have a self-dual $\mathbb{Z}_4$ 2-form symmetry (the so-called defect group, see for instance \cite{Morrison:2020ool}). This 2-form symmetry is known to participate in a 3-group structure with the 1-form symmetry of the form \cite{Apruzzi:2022dlm}: $\delta C = \mathcal{P} (B_s + B_c) + 2B_s\cup B_c$, where we use $C$ for the background field for the 2-form symmetry, $B_{s,c}$ for the background fields for the two generators of the $\mathbb{Z}_2 \times \mathbb{Z}_2$ 1-form symmetry and $\mathcal{P}(B)$ stands for the Pontryagin square of $B$.
\end{itemize} 

We next want to reduce this theory on a torus to 4d, and explore what happens to the discrete symmetries in the reduction. We shall approach this by reducing first to 5d and then to 4d. 

\subsection{Reduction to 5d}

We shall begin with the 5d reduction. When reduced on a circle, the theory is known to have an effective description as a 5d $SU(2)$ gauge theory\footnote{Here we assume a choice of polarization of the 6d SCFT such that the defect group reduces to a 1-form symmetry in 5d. It is possible to get the $SO(3)$ gauging by choosing a different polarization.} gauging the $SO(3)$ global symmetry of $4$ disconnected $E_1$ 5d SCFTs \cite{Hayashi:2017jze}, see figure \ref{SO8dr}. We remind the reader that the $E_1$ SCFT is a 5d SCFT that is the UV completion of an $su(2)$ gauge theory with $\theta=0$ \cite{Seiberg:1996bd}. We shall next need a few properties of this SCFT that we briefly summarize below:

\begin{figure}
\center
\includegraphics[width=0.3\textwidth]{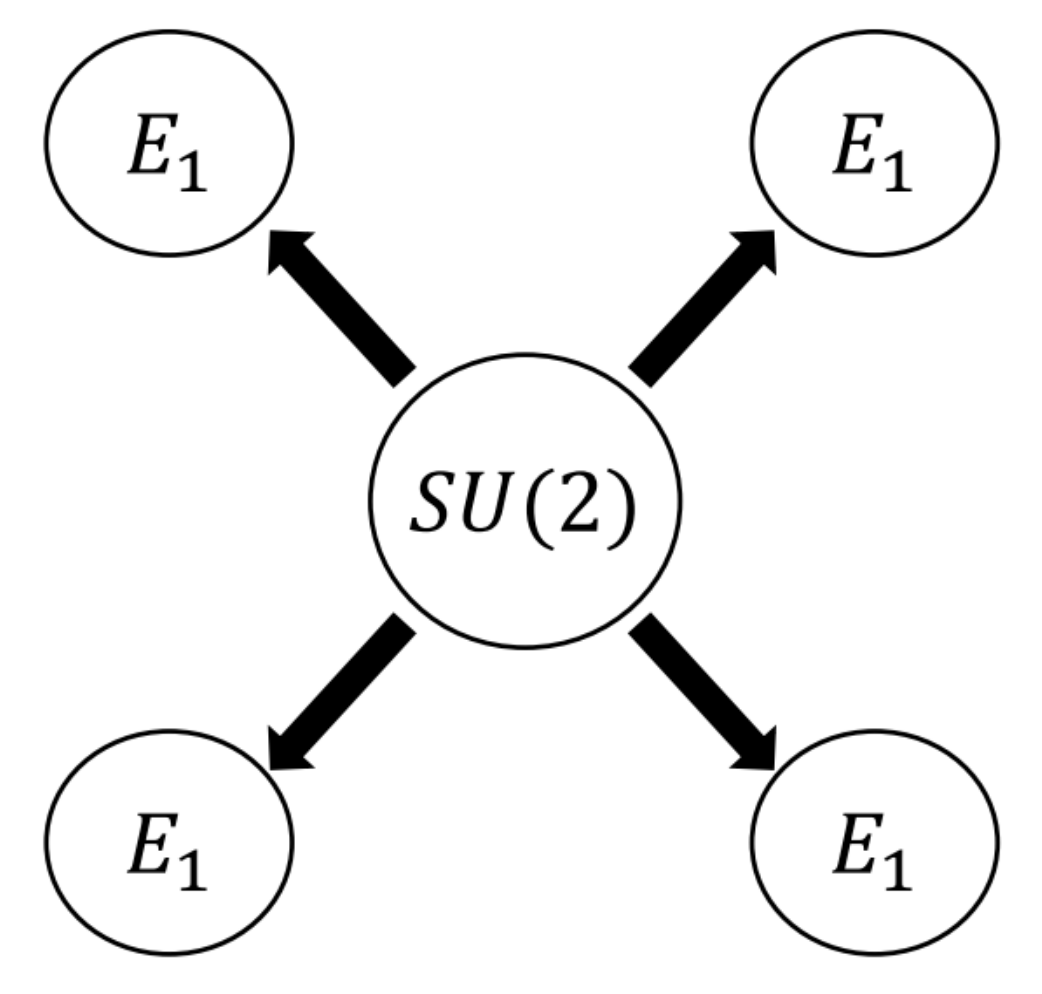} 
\caption{The 5d effective description of the 6d SCFT reduced on a circle. The central circle denotes an $SU(2)$ gauge theory, and the smaller corner circles denote $E_1$ SCFTs. The arrow denotes a gauging of the $SU(2)$ flavor symmetry of the $E_1$ SCFT by the $SU(2)$ gauge group.}
\label{SO8dr}
\end{figure}

\begin{itemize}
 \item The SCFT has an $SO(3)$ global symmetry \cite{Apruzzi:2021vcu}. A mass deformation associated with this global symmetry will initiate an RG flow leading to the pure $su(2)$ gauge theory. Note that in the above 5d theory, the $SO(3)$ is gauged so such a mass deformation is not possible (rather it is associated with the Coulomb branch of the gauge theory).
\item The SCFT has a $\mathbb{Z}_2$ 1-form symmetry. Here we assume a specific choice of charge lattice such that the low-energy gauge theory is $SU(2)$. The 1-form symmetry then appears in the low-energy theory as its electric 1-form symmetry.   
 \item The 1-form symmetry and $SO(3)$ global symmetry participate in a 2-group given by \cite{Apruzzi:2021vcu}: $\delta B = \beta w_2$, for $B$ the background gauge field for the 1-form symmetry, $w_2$ the second Stiefel-Whitney class of $SO(3)$ and $\beta$ the Bockstein homomorphism associated with the extension: $0 \rightarrow \mathbb{Z}_2 \rightarrow \mathbb{Z}_4 \rightarrow \mathbb{Z}_2 \rightarrow 0$.
 \item The two are also involved in a mixed anomaly \cite{BenettiGenolini:2020doj,Apruzzi:2021nmk}, given by the anomaly theory: $\frac{i\pi}{2}\int w_2 \cup \mathcal{P} (B)$, where the notation is as before.
\end{itemize}

We shall now explore the relation between the global symmetries of the 6d and 5d theories. We begin by listing the global symmetries of the 5d theory. First there is the 5d supersymmetry algebra (the theory is not conformal) and $su(2)$ R-symmetry. Additionally, we also have the $u(1)$ instantonic symmetry which is associated with the circle. We also have several discrete symmetries. First there is the $S_4$ symmetry associated with permuting the $4$ $E_1$ SCFTs. Second there are $5$ $\mathbb{Z}_2$ 1-form symmetries, $4$ coming from the 1-form symmetries of the $E_1$ SCFTs and one from the electric 1-form symmetry of the $SU(2)$ \footnote{Recall that the $E_1$ SCFT has an $SO(3)$ global symmetry so the center of the gauge symmetry should act trivially on local operators. Nevertheless, there is one subtly here involving the instantons of the $SU(2)$. Specifically, instanton operators associated with the $SU(2)$ gauge theory can be charged under its center. This happens for instance in a pure $SU(2)$ gauge theory with theta angle $\theta=\pi$, and is responsible for the fact that the $\tilde{E}_1$ SCFT has no 1-form symmetry. As we shall see, this 1-form symmetry is needed for the symmetry matching, suggesting that the $SU(2)$ theta angle here is $\theta=0$ so that the 1-form symmetry be consistent with the instanton operators.}.

The mapping between the continuous symmetries of the two theories is relatively understood, where the 5d supersymmetry algebra and R-symmetry come from the 6d superconformal symmetry, after the conformal part is broken, and the $u(1)$ is associated with the KK modes. We would next like to understand the relation between the discrete symmetries of the 5d theory to those of its 6d parent. From 6d, we would expect an $S_3$ global symmetry coming from the outer automorphism symmetry, a $\mathbb{Z}_2 \times \mathbb{Z}_2$ 0 and 1-form symmetries coming from the 6d 1-form symmetry, and a $\mathbb{Z}_4$ 1-form symmetry coming from the self-dual 2-form symmetry\footnote{As previously mentioned, the 6d defect group can reduce to either a 5d 1-form or 2-form symmetry depending on the polarization. Here we assume a choice was made giving the 1-form symmetry.}. We next want to understand the relation between these and the symmetries we observe in 5d.

We begin with the 0-form symmetries. From 6d we expect an $S_3$ and a $\mathbb{Z}_2 \times \mathbb{Z}_2$ global symmetry, while in 5d we observe an $S_4$ global symmetry. The two are actually related as $S_4$ is the semi-direct product of $S_3$ and $\mathbb{Z}_2 \times \mathbb{Z}_2$. In fact this follows from the following more general statement: the symmetries of the Dynkin diagram of an affine Lie-group $G$ are given by the extension of the outer automorphism symmetry of $G$ (which are the symmetries of the Dynkin diagram of the Lie-group $G$) by the center of $G$. More concretely, in our case we have $S_4 = (\mathbb{Z}_2 \times \mathbb{Z}_2) \rtimes S_3$, with the $S_3$ acting on the $\mathbb{Z}_2 \times \mathbb{Z}_2$ via its outer automorphism permuting its 3 $\mathbb{Z}_2$ subgroups. The 5d theory has the form of the Dynkin diagram of affine $so(8)$, and so inherits its $S_4$ symmetry. However, in 6d we had the $S_3$, coming from the outer automorphisms of standard $so(8)$, and the $\mathbb{Z}_2 \times \mathbb{Z}_2$ 1-form symmetry coming from the center of $so(8)$. Furthermore, the $S_3$ should have a non-trivial action on the $\mathbb{Z}_2 \times \mathbb{Z}_2$, acting like its outer automorphism. When reduced to 5d, this non-trivial action of the 0-form symmetry on the 1-form symmetry becomes an extension of the two 0-form symmetries, leading them to combine and form the $S_4$ symmetry observed in the 5d theory.        

This leads us to consider the 1-form symmetries. Here we expect a $\mathbb{Z}_2 \times \mathbb{Z}_2 \times \mathbb{Z}_4$ 1-form symmetry \cite{Morrison:2020ool}. However, in 5d we seem to only observe $\mathbb{Z}_2$ 1-form symmetries, so how are these related? Here is where the previously mentioned properties of the 5d $E_1$ SCFT, notably its 2-group structure, come into play. The important point is that once we gauge the $SO(3)$ global symmetry of an $E_1$ SCFT by an $SU(2)$ gauge group, the second Stiefel-Whitney class of its $SO(3)$ global symmetry becomes identified with the background gauge field for the $\mathbb{Z}_2$ 1-form symmetry of the gauge $SU(2)$. As such the 2-group should become a standard extension, and similarly the 't Hooft anomalies of the $E_1$ SCFTs would become 't Hooft anomalies connecting the various 1-form symmetries.

Next we examine how the extension happens in the case above. This is somewhat complicated by the fact that there are four $E_1$ SCFTs, and so five $\mathbb{Z}_2 $ 1-form symmetries. There is a similar interesting case where we gauge only two $E_1$ SCFTs (see appendix \ref{App:5dSU4}), where the resulting analysis is easier to follow. To perform the analysis, we find it convenient to introduce a mock symmetry TQFT. The symmetry TQFT is a dynamical (that is not necessarily invertible) topological quantum field theory that can be associated with a given theory and lives in one higher dimension. The symmetry TQFT can be used to keep track of the symmetry structure of a theory like its symmetries and their 't Hooft anomalies, as well as those of all theories related to it by topological manipulations. Generically, a specific theory is chosen by giving appropriate boundary conditions, with different boundary conditions being related by various topological manipulations, like gauging discrete symmetries. We refer the reader to \cite{Gaiotto:2020iye,Apruzzi:2021nmk,Freed:2022qnc,Kaidi:2022cpf} for more on symmetry TQFTs. For our purposes, we rely on the fact that the 2-group structure can be represented as a term in the action of the symmetry TQFT\footnote{This is essentially as gauging the 1-form symmetry participating in the 2-group converts it into a 't Hooft anomaly \cite{Tachikawa:2017gyf}.}, which allows us to study it by manipulating said action. Here we shall only use it as a computational tool, and shall only include those terms relevant to understanding the 2-group structure, hence why we call it a mock symmetry TQFT. For the case at hand, it takes the form:

\be \label{SymTQFT}
\pi i \int \hat{c} \cup \delta \hat{B} + c_1 \cup \delta B_1 + c_2 \cup \delta B_2 + c_3 \cup \delta B_3 + c_4 \cup \delta B_4 + c_1 \cup \beta \hat{B} + c_2 \cup \beta \hat{B} + c_3 \cup \beta \hat{B} + c_4 \cup \beta \hat{B} ,
\ee
    
Here $\hat{B}$ is the background field for the $SU(2)$ electric 1-form symmetry, $B_{1,2,3,4}$ are the background fields for the four $E_1$ SCFT's 1-form symmetries, and $\hat{c}, c_{1,2,3,4}$ are the conjugate 3-form gauge fields. The first five terms are the kinetic terms for the five $\mathbb{Z}_2$ 1-form symmetries. These couple them to the conjugate 3-forms ensuring that only one can receive Dirichlet boundary conditions, and as such describe a global symmetry. The other four terms are 't Hooft anomalies that lead to the 2-group when the $c$ fields are given Neumann boundary conditions. Here we have also ignored the terms associated with the 't Hooft anomalies of the $E_1$ SCFTs, which will be discussed separately later.

Next, we can rewrite \eqref{SymTQFT} as:

\bea 
& & \pi i \int \hat{c} \cup \delta \hat{B} + (c_1 + c_2 + c_3 + c_4) \cup \delta B_1 + c_2 \cup \delta (B_2 + B_1) + c_3 \cup \delta (B_3 + B_1) \nonumber \\ & & + c_4 \cup \delta (B_4 + B_1) + (c_1 + c_2 + c_3 + c_4) \cup \beta \hat{B} . \label{SymTQFTR}
\eea 

Now it is convenient to introduce the change of variables:

\bea \label{BFtrans}
\begin{pmatrix}
  c^+_1 \\
  c_2 \\
	c_3 \\
	c_4
\end{pmatrix} = \begin{pmatrix}
  1 & 1 & 1 & 1 \\
  0 & 1 & 0 & 0 \\
	0 & 0 & 1 & 0 \\
	0 & 0 & 0 & 1
\end{pmatrix}\begin{pmatrix}
  c_1 \\
  c_2 \\
	c_3 \\
	c_4
\end{pmatrix} \; , \; \begin{pmatrix}
  B_1 \\
  B^+_2 \\
	B^+_3 \\
	B^+_4 \\
\end{pmatrix} \rightarrow \begin{pmatrix}
 1 & 0 & 0 & 0 \\
 1 & 1 & 0 & 0 \\
 1 & 0 & 1 & 0 \\
 1 & 0 & 0 & 1
\end{pmatrix}\begin{pmatrix}
  B_1 \\
  B_2 \\
	B_3 \\
	B_4
\end{pmatrix} .
\eea

\eqref{SymTQFTR} can now be rewritten as:

\be
\pi i \int \hat{c} \cup \delta \hat{B} + c^+_1 \cup (\delta B_1 + \beta \hat{B} ) + c_2 \cup \delta B^+_2  + c_3 \cup \delta B^+_3 + c_4 \cup \delta B^+_4 .
\ee 

The above suggests that $\hat{B}$ and $B_1$ combine to form the $\mathbb{Z}_4$ gauge field: $X = \hat{B} + 2B_1$. Specifically, we have that $\delta X = 0 \rightarrow \delta B_1 = \beta \hat{B}$, which is implied by the equation of motion of $c^+_1$. All in all, we see that we have one $\mathbb{Z}_4$ and three $\mathbb{Z}_2$ 1-form symmetries, associated with the background gauge fields $X$, $B^+_{2,3,4}$. Here, the electric 1-form symmetry of the $SU(2)$ gauge group combines with the diagonal 1-form symmetry of the four $E_1$ SCFTs to form a $\mathbb{Z}_4$ 1-form symmetry\footnote{This may seem strange as $B_1$ is the background field appearing in $X$. However, in the symmetry TQFT the topological operators are built from the fields receiving Neumann boundary conditions, which here are the $c$ fields. $B_1$ couples to the diagonal combination of all the $c_i$'s suggesting its the diagonal combination that participates in the extension.}, which should map to the reduction of the self-dual 2-form symmetry of the 6d SCFT. Note that it is invariant under the $S_4$ as expected. 

This leaves us with three $\mathbb{Z}_2$ 1-form symmetries. Two of these should map to the $\mathbb{Z}_2 \times \mathbb{Z}_2$ 1-form symmetry of the 6d SCFT, while the third appears to be accidental. For instance, we can take the diagonal combination of the 1-form symmetries of the two upper $E_1$ SCFTs and another diagonal combination of the the two rightmost ones. These indeed form a $\mathbb{Z}_2 \times \mathbb{Z}_2$ 1-form symmetry under which the $S_3$ permuting the three $E_1$ SCFT (all save the bottom left one) acts by the permutation of its $\mathbb{Z}_2$ subgroups. The resulting structure is consistent with what we expect from the 6d SCFT on the circle.  

Overall, we see that we can indeed match the symmetries, though we seem to be forced to accept one of the $\mathbb{Z}_2$ 1-form symmetries as an accidental enhancement.

Finally, we wish to consider the effect of the 3-group structure in the 6d SCFT. When reduced on the circle, the 6d 3-group structure reduces as:

\be \label{3g5d}
\delta C = \mathcal{P} (B^{6d}_s + B^{6d}_c) + 2B^{6d}_s\cup B^{6d}_c \rightarrow \delta X = 2A_s\cup B^{5d}_s + 2A_c\cup B^{5d}_c,
\ee 
where $B^{5d}_{s,c}$ now describe the background gauge field for the 5d $\mathbb{Z}_2 \times \mathbb{Z}_2$ 1-form symmetries coming from the 6d one, while $A_{s,c}$ describe the background gauge field for the 5d $\mathbb{Z}_2 \times \mathbb{Z}_2$ 0-form symmetries coming from it. As before, we use $X$ for the background gauge field for the 5d $\mathbb{Z}_4$ 1-form symmetry coming from the 6d 2-form symmetry. 

We see then that the 6d 3-group structure reduces to a 2-group structure involving the 1-form and 0-form symmetries that originate from the higher form symmetries of the 6d SCFT. The compactification picture suggests that this structure should be present in the 5d theory. We shall not study this in-depth here, but we do note the following. The 5d 1-form symmetries coming from the 1-form symmetries of the 6d SCFT, transform non-trivially under the $S_4$ 0-form symmetry. The non-trivial transformation under the $S_3$ part is expected from the 6d SCFT, but usually we would expect the 1-form symmetry to commute with itself wrapping the circle. However, here the two appear to not completely commute. It is tempting to attribute this failure of commutativity to the 2-group relation \eqref{3g5d}. This comes about since the analogous version involving only 0-form symmetries leads precisely to such non-commutativity, as we shall see when discussing the 4d reduction.    

\subsubsection*{'t Hooft anomalies}

Finally, the 5d theory also possesses several 't Hooft anomalies originating from the known anomalies of the $E_1$ SCFT and $SU(2)$ gauge theory, see \cite{BenettiGenolini:2020doj,Apruzzi:2021nmk}. There are two main sources of anomalies. One is the $SU(2)$ gauge theory, while the other is the anomaly of the $E_1$ SCFT we previously mentioned. Here we shall briefly mention them for completeness.

The $SU(2)$ gauge theory leads to a mixed anomaly involving its 1-form and instantonic $U(1)$ symmetry. The form of the anomaly is:

\be
\frac{\pi i}{2}\int \tilde{C}_1 (U(1)) \cup \mathcal{P}(\hat{B}) ,
\ee 
where here $\tilde{C}_1 (U(1))$ stands for the mod 4 reduction of the first Chern class of the $U(1)$, while $\mathcal{P}(\hat{B})$ is the Pontryagin square of $\hat{B}$. Note that we can rewrite this using the field $X$ as:

\be
\frac{\pi i}{2}\int \tilde{C}_1 (U(1)) \cup \mathcal{P}(X) ,
\ee
with no need for additional anomaly terms.

The second source of anomalies come from the anomaly in the $E_1$ SCFT. As we noted, each such SCFT is expected to also have an anomaly involving its 1-form and $SO(3)$ global symmetry of the form: $\frac{\pi i}{2} \int w_2 (SO(3))\cup \mathcal{P}(B)$. In our case, due to the gauging, we identify $w_2 (SO(3)) = \hat{B}$ for all four $E_1$ SCFTs. This leads to the anomaly term:

\be
\frac{\pi i}{2} \int \hat{B}\cup (\mathcal{P}(B_1) + \mathcal{P}(B_2) + \mathcal{P}(B_3) + \mathcal{P}(B_4)) .
\ee

Note that here it is quite natural to extend the $\mathbb{Z}_2$ valued $\hat{B}$ to the $\mathbb{Z}_4$ valued $X$:

\be \label{SO85dAnom}
\frac{\pi i}{2} \int X\cup (\mathcal{P}(B_1) + \mathcal{P}(B_2) + \mathcal{P}(B_3) + \mathcal{P}(B_4)) ,
\ee
where we do not need to add any additional terms as the Pontryagin square is even on spin manifolds.

It would be interesting to better understand the anomalies of the 5d theory and compare with those expected from the 6d SCFT on a circle, though we shall not pursue this here.

\subsection{Reduction to 4d}
\label{SO84dR}

We can next consider the reduction to 4d. This can be done by reducing the 5d theory on the circle. The $SU(2)$ gauging should just reduce to an analogous gauging in 4d, as the gauge coupling is IR free in 5d. The main issue is understanding what does the $E_1$ SCFT reduces to. Fortunately, the results of \cite{Ganor:1996pc} suggest that the low-energy theory is an $\mathcal{N}=2$ $U(1)$ gauge theory with two massless hypermultiplets. Here, the $SO(3)$ flavor symmetry of the $E_1$ SCFT should map to the symmetry rotating the two hypers. 

Combining all this, we see that the resulting 4d theory should be a $D_4$ shaped quiver with an $SU(2)$ central node surrounded by 4 $U(1)$ nodes. Next, we want to understand the mapping of the symmetries between the 5d and 4d theories. Here it is convenient to first discuss the matching between the $E_1$ SCFT and the 4d $U(1)$ theory, and then employ this to study the full case.

\subsubsection*{4d reduction of the $E_1$ SCFT}

We begin with the circle reduction of the 5d $E_1$ SCFT. This was worked out in \cite{Ganor:1996pc} to be a 4d $\mathcal{N}=2$ $U(1)$ gauge theory with two hypermultiplets. We shall next consider the mapping of symmetries between 5d and 4d, which will be useful later when we discuss the 4d reduction of the 6d $so(8)$ SCFT.

As previously mentioned, the 5d SCFT has an $SO(3)$ 0-form global symmetry, as well as a $\mathbb{Z}_2$ 1-form symmetry. The two are also connected via a 2-group structure, and a mixed anomaly. We would like to match this structure with the 4d theory. The $SO(3)$ global symmetry naturally matches the symmetry rotating the two hypermultiplets, where the central element is quotiented as it is equivalent to a gauge transformation. There are two natural options to match the $\mathbb{Z}_2$ 1-form symmetry. One is to embed it in the $U(1)$ magnetic 1-form symmetry. Another option is to have it as an electric 1-form symmetry. Essentially, we take the hypers to have charge 2, and add a Wilson line with charge 1. The two options are related by gauging.

We next want to determine which option we should get by matching with the properties we expect from 5d. One property is the existence of the 2-group structure: $\delta B = \beta w_2$. This should reduce to a similar structure in 4d. Indeed it is known that such a structure can arise in these types of systems. Specifically, consider the case where we take the hypers to have charge 2, and we have an electric 1-form symmetry. This 1-form symmetry acts on the Wilson line with charge 1. Note that if the hypers had charge 4, this 1-form symmetry would be the subgroup of the $\mathbb{Z}_4$ 1-form symmetry acting on the Wilson lines with charges $1-3$. Coupling the theory to a non-trivial $w_2$ necessitates the introduction of a magnetic flux acting on the charge 2 hypers. Such flux corresponds to the generator of the $\mathbb{Z}_4$ 1-form symmetry, and as such the two would be related by the 2-group: $\delta B = \beta w_2$ (this basically follows from the closeness of the $\mathbb{Z}_4$ 1-form symmetry connection $X = w_2 + 2B$). As such, we see that the frame with the electric $\mathbb{Z}_2$ 1-form symmetry can accommodate the presence of the 2-group.

Next we turn to consider the anomaly. In 5d, it takes the form: $\frac{\pi i}{2} \int w_2 \cup \mathcal{P}(B)$. When reduced on the circle it gives: $\pi i \int w_2 \cup B \cup A$, where $A$ is the background connection for the $\mathbb{Z}_2$ 0-form symmetry coming from the 1-form symmetry wrapping the circle (here $\int_{S^1} w_2 = 0$, so the 2-group relation trivializes for $A$). This raises the question of what this 0-form symmetry is in the 4d theory. Note that the 4d theory possess a new symmetry not present in the 5d SCFT: charge conjugation. Specifically, the 5d SCFT appears to be real, as all global symmetries and the 5d gauge theory description are both real. However, the 4d theory contains complex representations, leading to the presence of a new symmetry. We shall next argue that one can match the properties of the 4d and 5d theories, both for the $E_1$ SCFT and the circle reduction of the 6d $so(8)$ SCFT, if we identify charge conjugation as the 1-form symmetry on the circle. 

Specifically, we would like to show the presence of the anomaly $\pi i \int w_2 \cup B \cup A$. Actually, like the 2-group structure, this anomaly is known to naturally arise in $so(4k+2)$ gauge theories with vector matter, where the case we consider here is just the $k=0$ case. This works as follows. Consider a pure $Spin(4k+2)$ gauge theory. We have a $\mathbb{Z}_4$ 1-form symmetry\footnote{For $k=0$ we have a full $U(1)$, though we can reduce it to $\mathbb{Z}_4$ by including charge $4$ matter as done above.} and a $\mathbb{Z}_2$ 0-form symmetry associated with charge conjugation (outer automorphism of $Spin(4k+2)$). We can think of the 1-form symmetry as being made from two $\mathbb{Z}_2$ 1-form symmetries via an extension, in which case the three symmetries would be connected by the  following relation: $\delta B_H = \beta B_Q + A \cup B_Q$ \cite{Hsin:2020nts}, where we use $B_H$, $B_Q$ for the background connections of the subgroup and quotient $\mathbb{Z}_2$ 1-form symmetries, respectively \footnote{This essentially follows as charge conjugation does not commute with the $\mathbb{Z}_4$. If all symmetries where 0-form, they would combine to form the dihedral group $D_4$, and we would have had the following relation: $\delta A_H = \beta A_Q + A \cup A_Q$ (see appendix \ref{App:D4}). Replacing the 1-forms $A_H$ and $A_Q$ by the 2-forms $B_H$ and $B_Q$ for the 1-form symmetries, gives the desired relation.}. 

Adding an even number of vector matter amounts to setting $B_Q=w_2$, and the relation becomes: $\delta B_H = \beta w_2 + A \cup w_2$. Note that this gives us the 2-group relation we had before plus an additional term involving the charge conjugation symmetry. Additionally, gauging $B_H$, so that we transform to the $SO$ variant, gives the anomaly: $\pi i \int w_2 \cup \tilde{B}_H \cup A$, for $\tilde{B}_H$ the connection for the dual 1-form symmetry. Thus, we can naturally get both the 2-group and the 1-form symmetry, but we seem to need different global structures for each case. Specifically, the 2-group seem to be present for the case of $Spin(2)$, while the anomaly requires the $SO(2)$ global structure. 

The solution to this issue seems to be the observation in \cite{Lee:2021crt} that quantum corrections, particularly the presence of fermionic zero modes in the monopole background, may affect the symmetry structure. Specifically, it was argued that for an $\mathcal{N}=1$ $so(2n)$ gauge theory with $2n_f$ vector chirals, the question of whether you get the 2-group structure, $\delta B_H = \beta w_2$, or the anomaly, $\pi i \int B_H \cup \beta w_2$, may differ from the results in \cite{Hsin:2020nts} due to the presence of fermionic zero mode. Particularly, in the case where both $n$ and $n_f$ are odd, the $SO$ variant should actually have the 2-group structure: $\delta B_H = \beta w_2$. The reference did not study the case of the 2-group $\delta B_H = A \cup w_2$, and the corresponding anomaly, but it seems reasonable that this remains as before\footnote{One issue with the 2-group and anomaly involving $\beta w_2$ is that they are only present for $n$ odd. Given an $\mathcal{N}=1$ $so(2n)$ gauge theory with $2n_f$ vector chirals, we can consider its Seiberg dual, which is an $so(2n')$ gauge theory with $2n_f$ vector chirals, where $n'=n_f-n+2$. Note that if both $n$ and $n_f$ are odd, then $n'$ is even, and we cannot match the generalized symmetry structure in this case. The corrections provided by the zero modes ensures that this structure would match. This problem does not seem to exist for the 2-group and anomaly involving $A \cup w_2$, as these are present for every $n$.}. Note that the case considered here corresponds to $n=n_f=1$, that is the case where both are odd, and the results above suggests that the $SO(2)_+$ variant has both the 2-group, $\delta B_H = \beta w_2$, and the anomaly, $\pi i \int w_2 \cup B_H \cup A$. As such we conclude that this is the desired variant.

To summarize, we have seen that the 5d $E_1$ SCFT reduces to a 4d $\mathcal{N}=2$ $U(1)$ gauge theory with 2 charge $1$ hypermultiplets. The $\mathbb{Z}_2$ 1-form symmetry reduces to the 1-form symmetry of the $U(1)$ theory (for the $SO$ variant in question this would be the mod 2 reduction of the monopole charge, so there appears to be an accidental enhancement of the 1-form symmetry to a $U(1)$). Additionally, it also reduces to a $\mathbb{Z}_2$ 0-form symmetry which we identify with charge conjugation. Finally, the $SO(3)$ flavor symmetry just reduces to the $SU(2)/\mathbb{Z}_2$ symmetry rotating the 2 hypers. The anomaly and 2-group structure appear to match between the two descriptions. Next we use this to match the symmetries between the 4d theory and the 6d SCFT reduced on a torus.

\subsubsection*{4d reduction of the full 5d theory}

We can now consider the reduction of the complete 5d theory describing the 6d SCFT reduced on a circle. As we have previously seen, this 5d theory is just an $SU(2)$ gauging of 4 $E_1$ SCFTs. When reduced to 4d, each $E_1$ SCFT should reduce to a $U(1)$ gauge theory with 2 charge $1$ hypermultiplets, while the $SU(2)$ gauging would reduce to the analogous gauging in 4d as it is IR free in 5d. As such, we end up with an affine $D_4$ shaped quiver with a central $SU(2)$ node surrounded by four $U(1)$ gauge nodes. We next discuss the matching of properties between the 4d theory and the 5d theory or the 6d SCFT.

 \paragraph*{Moduli space} The 4d theory possesses a 5 dimensional Coulomb branch and a 1d Higgs branch whose geometry is that of $\mathbb{C}^2/\hat{D}_4$. We have noted that the original 6d SCFT has the same Higgs branch, as should be since the Higgs branch should remain the same under the $T^2$ reduction. The Coulomb branch geometry can change, but we can still match its dimension. In the 6d SCFT we expect 4 directions to come from the $so(8)$ vectors when reduced on the cycles of the torus, as well as 1 more coming from the tensor. This matches the observed dimension of the 4d Coulomb branch.
   
\paragraph*{Conformal manifold} While the $U(1)$ gauge part is IR free, the $SU(2)$ gauge group sees $4$ doublet hypermultiplets, and as such, is conformal. This suggests the presence of a 1d "conformal" manifold spanned by the $SU(2)$ coupling constant. Note that this coupling constant originates from the 5d coupling, which in turn should be proportional to the inverse of the radius of the 6d circle. This suggests that $g^2_{4d} \sim \frac{R_1}{R_2}$, and as such that the value of the coupling constant should be related to the imaginary part of the modular variable determining the complex structure of the torus. Therefore, we identify this 1d "conformal" manifold as coming from the choice of complex structure for the torus, exactly as in $\mathcal{N}=4$ SYM. Furthermore, it is known that $SU(2)$ with 4 flavors has an $SL(2,\mathbb{Z})$ duality symmetry as expected, see \cite{Seiberg:1994aj}\footnote{In class S, this symmetry is usually attributed to the mapping class group of the 4-punctured sphere, which happens to be the same as that of the torus. This construction provides a realization of it in terms of the mapping class group of the torus.}.
  
 \paragraph*{1-form symmetries} We expect to get a $\mathbb{Z}_2 \times \mathbb{Z}_2 \times \mathbb{Z}_4$ 1-form symmetry from the higher dimensional construction. In the 5d theory, this is apparent as it is just the 1-form symmetries we have in the 5d theory, and we don't observe any symmetries of higher form degree. In the 6d SCFT, the $\mathbb{Z}_2 \times \mathbb{Z}_2$ part comes from the 1-form symmetry of the 6d SCFT, while the $\mathbb{Z}_4$ comes from the self-dual 2-form symmetry\footnote{Because it is self-dual, we only get one 1-form symmetry rather than two from the two cycles.}.
 In the 4d theory, we observe a $\mathbb{Z}_2$ 1-form symmetry coming from the $SU(2)$ gauge group, as well as four magnetic $U(1)$ 1-form symmetries associated with the four $U(1)$ gauge groups. Following our discussion about the reduction of the $E_1$ SCFT, we identify the mod 2 part of the $U(1)$ 1-form symmetries with the $\mathbb{Z}_2$ 1-form symmetries coming from the four $E_1$ SCFTs. Due to the presence of the 2-group, the diagonal combination of these, and the $\mathbb{Z}_2$ 1-form symmetry coming from the $SU(2)$ gauge group, combine to form a $\mathbb{Z}_4$ 1-form symmetry\footnote{There is a simpler story if we chose the $Spin(2)$ variants, that is we regard the four edge theories as $U(1)$ gauge theories with charge $2$ hypermultiplets. In this case we have electric $\mathbb{Z}_2$ 1-form symmetries, obeying the same 2-group relation, and we again have the enhancement to $\mathbb{Z}_4$. This enhancement can now be seen from the field theory as the $\mathbb{Z}_2$ center we need to quotient by is the diagonal between the $SU(2)$ and the four $U(1)$ groups, and so combines with the diagonal $\mathbb{Z}_2$ electric 1-form symmetry.}, which should match the $\mathbb{Z}_4$ part of the 1-form symmetry expected from 6d. Two of the remaining $\mathbb{Z}_2$ symmetries then match the $\mathbb{Z}_2 \times \mathbb{Z}_2$ part, with the last one being identified as an accidental symmetry (here we have further accidental symmetries as we have $U(1)$ magnetic 1-form symmetries).
 
 \paragraph*{0-form symmetries} There are many 0-form symmetries that we can try to match. We begin with what we do not have, which is continuous 0-form flavor symmetries. Indeed, there are no continuous symmetries in the 6d SCFT, save for the superconformal symmetry, and we indeed observe none also in the 4d theory, save for the supersymmetry algebra (conformal symmetry was broken by the reduction). We do however expect to get many discrete 0-form symmetries. From the 6d perspective, these come from the 0-form, 1-form and self-dual 2-form symmetries of the 6d SCFT. Specifically, we expect an $S_3$ 0-form symmetry from the 0-form symmetries of the SCFT, a $\mathbb{Z}^4_2$ symmetry from the $\mathbb{Z}_2 \times \mathbb{Z}_2$ 1-form symmetry of the 6d SCFT wrapping the two cycles of the torus, and a $\mathbb{Z}_4$ 0-form symmetry from the self-dual 2-form symmetry wrapping the full torus. We shall first discuss the $S_3$ and $\mathbb{Z}^4_2$ part, which combine to form one large symmetry group due to the non-trivial action of the $S_3$ on the $\mathbb{Z}^4_2$ part.

 It is convenient here to match the structure between the 4d and 5d theories first, and then discuss the 6d interpretation. We have seen that in 5d we have the $S_4 = (\mathbb{Z}_2 \times \mathbb{Z}_2) \rtimes S_3$ symmetry coming from the $S_3$ action on the $\mathbb{Z}^2_2$ part wrapping the circle. We have also concluded that the $\mathbb{Z}_2$ 1-form symmetry of the $E_1$ SCFT should reduce to charge conjugating the $U(1)$. As such we expect the 4d 0-form symmetry to be built from the $S_4$ rotating the four $U(1)$ nodes, as well as the charge conjugation of each individual node. This would indeed give the right structure, as the $S_3$ would have a natural action on the charge conjugation symmetries.
 
There is a neat way to argue and identify what symmetry we have. The idea is to ungauge the $U(1)$ nodes. We then just have an $SU(2)$ gauge theory with 4 flavors. This theory is known to have an $so(8)$ flavor symmetry. Now we gauge its Cartan subalgebra, what symmetry do we have left? The general answer in this case is that we get the normalizer\footnote{The normalizer of $H$ in $G$ is the set of all elements of $g\in G$ such that $g^{-1} h g \in H$ for all $h\in H$. Note that when $H$ is a normal subgroup, the normalizer is $G$ and the left over symmetry is the quotient group, $\frac{G}{H}$, as expected. The above allows us to extend this to non-normal subgroups, in which case the normalizer is the largest subgroup of $G$ for which $H$ is a normal subgroup. A closely related definition is that of a centralizer, or commutant, that is the set of all $g$ for which $g^{-1} h g = h$ for all $h$. The centralizer is always a normal subgroup of the normalizer. The quotient contains the elements of the normalizer that act non-trivially on $H$. It is sometimes also said that the left over group after gauging is the centralizer, especially when gauging continuous symmetries. The subtle issue is that the centralizer is always a left over symmetry, but the part of the normalizer that is not in the centralizer may not be. This is as said part has a non-trivial action on $H$, which may not be a symmetry if one adds matter charged under $H$ that is not invariant under the transformation. For instance, in our case we gauge $U(1)^4$ without adding any matter, and so get the normalizer, but we can easily break the $S_4$ or the charge conjugation symmetry by adding matter not invariant under them. As such, it seems that the maximal left over symmetry is determined by the normalizer, and the minimal one by the centralizer, depending on the charged matter under the gauged symmetry.} of $u(1)^4$ in $so(8)$ qoutiented by the $u(1)^4$, see \cite{Cordova:2017kue} for instance. This is known to be the Weyl group of $so(8)$. As such the 4d theory actually possesses a discrete $W(D_4)$ global symmetry. Note that this symmetry is the semi-direct product of $S_4$, manifesting as the symmetry permuting the 4 $U(1)$ groups, with the $\mathbb{Z}_2$ groups acting as charge conjugation on each pair of $U(1)$ groups.

The group $W(D_4)$ contains a $\mathbb{Z}_2$ central element, corresponding to the charge conjugation of all four $U(1)$ groups (the entire 4d theory essentially). We can then quotient by it, and as shown in appendix \ref{App:WD4}, $W(D_4)/\mathbb{Z}_2 = (\mathbb{Z}_2 \times \mathbb{Z}_2)^2 \rtimes S_3$, where the $S_3$ acts by the simultaneous permutation of the 3 $\mathbb{Z}_2$ subgroups of both pairs of $\mathbb{Z}_2 \times \mathbb{Z}_2$. This is exactly the symmetry we expect from both the 5d description and the 6d SCFT due to the action of the $S_3$ 0-form symmetry on its $\mathbb{Z}_2 \times \mathbb{Z}_2$ 1-form symmetry.

This leaves us with the $\mathbb{Z}_4$ part. In 5d, the 1-form symmetry was made from the diagonal 1-form symmetry of the 4 $E_1$ SCFTs, and the $\mathbb{Z}_2$ 1-form symmetry coming from the center of the $SU(2)$. The former should correspond to the charge conjugation of all four $U(1)$ groups, which is the element we quotient out of $W(D_4)$. We see then that it actually fits as part of the $\mathbb{Z}_4$. However, for the $SU(2)$ part we do not observe the $\mathbb{Z}_2$ 0-form symmetry, but rather only the 1-form symmetry. This follows the usual observation where in Lagrangian theories the electric 1-form symmetry reduces only to an electric 1-form symmetry, while the magnetic $d-3$ form symmetry reduces only to a magnetic $d-4$ form symmetry. In \cite{Nardoni:2024sos}, the lose of the electric 0-form symmetry expected from the 1-form wrapping the circle was attributed to a spontaneous breaking due to the choice of origin of the scalar associated with the vector on the circle. We note, that here the $\mathbb{Z}_4$ is broken to its $\mathbb{Z}_2$ subgroup, which is consistent with this interpretation.

Overall, we see that we can match the symmetry structure between the 6d and 4d theories, up to the usual subtle issues including accidental symmetries and spontaneous breaking of symmetries. Nevertheless there is one more subtle issue that we would like to address\footnote{There is another subtle issue of whether we get the individual charge conjugation of each $U(1)$ group. The Weyl group of $so(8)$ only contains the conjugation of every pair. If the symmetry group is $o(8)$ then we would get the individual charge conjugation of each $U(1)$ group, and the symmetry is extended to the Weyl group of $usp(8)$. This is not important for the symmetry matching so we won't address it here.}. Specifically, the group $W(D_4)$ is not a direct product of $W(D_4)/\mathbb{Z}_2$ and the $\mathbb{Z}_2$, but rather a central extension, so how can we understand this from the 6d picture? This is where the 3-group structure plays a role. Recall that the 6d SCFT has a 3-group structure involving its 2-form and 1-form symmetries. When reduced on a circle this leads to a 2-group structure described by \eqref{3g5d}. We can further reduce on an additional circle to 4d and get:

\bea 
&& \delta X = 2A^{5d}_s\cup B^{5d}_s + 2A^{5d}_c\cup B^{5d}_c \rightarrow \delta A_X = 2A^{4d}_s\cup \hat{A}^{4d}_s + 2A^{4d}_s\cup \hat{A}^{4d}_s \rightarrow \nonumber \\ & & \delta A_H = A^{4d}_s\cup \hat{A}^{4d}_s + A^{4d}_s\cup \hat{A}^{4d}_s, \label{3g4d}
\eea  
where we use $A_X$ for the background connection of the $4d$ $\mathbb{Z}_4$ 0-form symmetry coming from the 6d 2-form one wrapping the torus, and $A^{4d}_{s,c}$, $\hat{A}^{4d}_{s,c}$ for the background connections of the $4d$ $(\mathbb{Z}_2 \times \mathbb{Z}_2)^2$ 0-form symmetry coming from the 6d 1-form symmetry wrapping the two cycles of the torus. In the last step we used $A_X = A_Q + 2 A_H$, to write $A_X$ in terms of the background fields of its $\mathbb{Z}_2$ subgroup and quotient, and dropped $A_Q$ as it acts trivially in the $4d$ limit. This is the reduction of the 5d 2-group relation on the circle, where there is also the direct reduction which we shall consider momentarily.

The term in \eqref{3g4d} describes a central extension of the $\mathbb{Z}^4_2$, spanned by the background fields $A^{4d}_{s,c}$ and $\hat{A}^{4d}_{s,c}$, by the $\mathbb{Z}_2$ symmetry spanned by $A_H$. Recall that the former are part of $W(D_4)/\mathbb{Z}_2$, while the latter is the simultaneous charge conjugation of all four $U(1)$ groups. As argued in appendix \ref{App:WD4} then, the relation  \eqref{3g4d} precisely gives the extension of $W(D_4)/\mathbb{Z}_2$ to $W(D_4)$ with the central element being the one described by $A_H$. As such, the entirety of the $W(D_4)$ global symmetry of the 4d theory originates from symmetries the 6d SCFT.   

Finally we note that the 5d 2-group should also lead to an analogous 4d 2-group. As discussed in appendix \ref{App:WD4}, the effect of \eqref{3g4d} is that some elements of $\mathbb{Z}^4_2$ commute only up to the $\mathbb{Z}_2$ spanned by $A_H$ (and as such commute in the quotient by not in the full group). We note that similar phenomena occurs between the 1-form symmetries and the $S_4$ permutation symmetry. Here the only one that commutes with the entire $S_4$ is the diagonal 1-form symmetry of all the $U(1)$ groups, which is the $\mathbb{Z}_2$ subgroup of the $\mathbb{Z}_4$ 1-form symmetry. In particular, there is no element that only transforms under the $S_3$ quotient. This seems at odds with the expectation that the 1-form symmetry commutes with their circle reductions. However, if we relax the requirement to commuting only up to the diagonal 1-form symmetry of all the $U(1)$ groups, then it is possible to find elements with the right properties, notably the diagonal 1-form symmetries under two $U(1)$ groups. It is therefore tempting to attribute this lack of commutativity to the 2-group structure. We expect the same to hold also for the 5d 2-group, though again we shall not pursue this further here.           

 \paragraph*{Anomalies} We can also try to match the anomalies between the 5d and 4d theories. Here the anomalies coming from the $E_1$ SCFTs should match based on the previous discussion regarding the 4d reduction of the $E_1$ SCFT. It would be interesting if this can be matched with a reduction of a 6d anomaly. This leaves the anomaly coming from the $SU(2)$ gauge theory, but it involves both the topological $U(1)$ symmetry and the 1-form symmetry which we do not observe in the 4d theory.

Overall, we see that we can match the symmetry structure of the 4d theory against that expected from the 6d SCFT. Of particular interest is the way the discrete 0-form symmetries match. Specifically, in the 6d theory we only have an $S_3$ 0-form symmetry, but also a $\mathbb{Z}^2_2$ 1-form symmetry on which the $S_3$ has a non-trivial action, and a 2-form symmetry involved in a 3-group structure with the 1-form symmetry. As we go down in dimensions, the 1-form symmetry on the cycles of the compact surface gives additional 0-form symmetries, and due to the non-trivial action of the $S_3$, they combine to form a larger group, $S_4$ in the 5d case and $W(D_4)/\mathbb{Z}_2$ in the 4d case. Furthermore, in 4d the 2-form symmetry wrapping the torus gives an additional 0-form symmetry, which due to the 3-group structure combines to enhance the group to $W(D_4)$. This provides an example of a case where a higher group structure, both split and non-split, leads to a semi-direct product in the split case and a central extension in the non-split case upon dimensional reduction.  

\section{Dimensional reduction for the general case}
\label{sec:soN}

Next, we consider the case of general $N$. As before there is a 1d tensor branch, on a generic point of which the theory has an effective description in terms of a 6d gauge theory, now $Spin(N)$ with $N-8$ vector hypers. We begin by noting several properties of these $6d$ SCFTs:

\begin{itemize}
 \item As mentioned, the 6d SCFT has a 1d tensor branch on a generic point of which the low-energy description is given by a $Spin(N)+(N-8)V$ gauge theory. The SCFT also has a $\frac{1}{2}(N-7)(N-8) + 1$ dimensional Higgs branch (see \cite{Akhond:2024nyr} for more information on the geometry of the Higgs branch). 
 \item In addition to the 6d superconformal symmetry, it also has a continuous $usp(2N-16)$ global symmetry ($PUSp(2N-16)$ for $N$ even and $USp(2N-16)$ for $N$ odd). These should be the only continuous symmetries of the SCFT. Like before, the anomalies under this symmetry can be encoded in an anomaly polynomial, which can be computed using the methods in \cite{Ohmori:2014kda}.
 \item For $N$ even, the SCFT should possess a discrete global symmetry, $\mathbb{Z}_2$, associated with the outer automorphism of $Spin(N)$.
 \item It has a discrete $\mathbb{Z}_2$ 1-form symmetry, associated with the central element of $Spin(N)$ that acts non-trivially only on the spinors. Again, this symmetry is thought to originate from the 6d SCFT (see for instance \cite{Apruzzi:2020zot,Apruzzi:2022dlm}). Although the 1-form symmetry is $\mathbb{Z}_2$ for all $N>8$, there are some differences between $N$ odd, $N$ even and divisible by $4$ and $N$ even but not divisible by $4$. 

Specifically, for $N$ odd, the full center of $Spin(N)$ is $\mathbb{Z}_2$ which is then the 1-form symmetry. For $N$ even however, the center is bigger, and $\mathbb{Z}_2$ is merely the subgroup of the center acting trivially on the vector. The part that acts non-trivially on the vector does not lead to a 1-form symmetry but does have interesting effects on the SCFT. This comes about as its would-be background field gets identified with the Stiefel Whitney class of the $PUSp(2N-16)$ global symmetry. Specifically, when $N=4m+2$, the center is $\mathbb{Z}_4$, the $\mathbb{Z}_2$ 1-form symmetry is the subgroup that acts trivially on the vector and $w_2(PUSp(2N-16))$ become identified with the quotient under it. This leads to a 2-group structure between the two of the form: $\delta B = \beta w_2(PUSp(2N-16))$, where $B$ denotes the background field for the $\mathbb{Z}_2$ 1-form symmetry. Similarly, when $N=4m$, the center is $\mathbb{Z}_2 \times \mathbb{Z}_2$, and the $\mathbb{Z}_2$ 1-form symmetry is again the subgroup that acts trivially on the vector. However, now since the center is a direct product of the two $\mathbb{Z}_2$ groups, there is no 2-group. Furthermore, for any even $N=2n$, there should be an additional 2-group involving also the parity symmetry which takes the form \cite{Hsin:2020nts}: $\delta B = A_P \cup w_2(PUSp(2N-16))$, where we use $A_P$ for the background field for the outer automorphism symmetry. The full 2-group can then be written as: $\delta B = n \beta w_2(PUSp(2N-16)) + A_P \cup w_2(PUSp(2N-16))$, where all coefficients should be taken mod 2.
\item Finally, as before, the SCFT should also have a self-dual $\mathbb{Z}_4$ 2-form symmetry (the so-called defect group). It participates in a 3-group structure with the 1-form symmetry of the form \cite{Apruzzi:2022dlm}: $\delta C = \mathcal{P} (B)$, where we again use $C$ for the background field for the 2-form symmetry and $B$ for the 1-form symmetry.
\end{itemize}

As can be seen from the above, one interesting aspect of this family of SCFTs for $N>8$ is that they have a continuous flavor symmetry, which for $N$ even, is not simply connected. Furthermore, this flavor symmetry, specifically its associated Stiefel-Whitney class, mixes with the 1-form symmetry through a 2-group structure. It is then interesting what happens to the 2-group structure if the 6d SCFT is compactified to 4d in the presence of a non-trivial Stiefel-Whitney class for the flavor symmetry. Such compactifications on tori have been studied for other families of 6d SCFTs and shown to lead to interesting 4d $\mathcal{N}=2$ SCFTs \cite{Ohmori:2018ona,Giacomelli:2020jel,Giacomelli:2020gee,Heckman:2022suy,Giacomelli:2024dbd}. Here we shall consider a similar reduction for the family of 6d SCFTs described above. As we shall see, this leads to 4d $\mathcal{N}=2$ non-conformal theories, similarly to the $N=8$ case, and so is not that interesting for studying 4d $\mathcal{N}=2$ SCFTs. However, it exhibits interesting discrete symmetry structures, that as we shall see, relate in a non-trivial way to the generalized symmetry structure of the 6d SCFT.       

Before discussing the Stiefel-Whitney reduction, it is convenient to first discuss the standard circle reduction of the 6d SCFTs. This would allow us to set the stage for the Stiefel-Whitney reduction, and is also of interest in its own right. As such, we shall begin with considering the circle reduction of this family of 6d SCFTs, and the mapping of symmetries between the 6d and 5d theories.  

\subsection{Standard circle reduction}

The reduction can be analyzed by utilizing a brane realization of the 6d SCFT \cite{Brunner:1997gk,Hanany:1997gh}. Specifically, we can engineer the 6d SCFT as the low energy theory on $N$ D6-branes, immersed in an O$6^-$-plane, and suspended between two NS5-branes that are stuck on the O$6$-plane. This leads to an $so(N)$ gauge theory living on the D6-brane. Conservation of the D6-brane charge then necessitates the presence of $N-8$ extended D6-branes branes on each side, which provide the vector matter. Overall, the low-energy theory living on the D6-branes, is a 6d $so(N)+(N-8)V$ gauge theory. Merging the two NS5-branes corresponds to going to the origin of the tensor branch, resulting in the 6d SCFT.

\begin{figure}
\center
\includegraphics[width=0.65\textwidth]{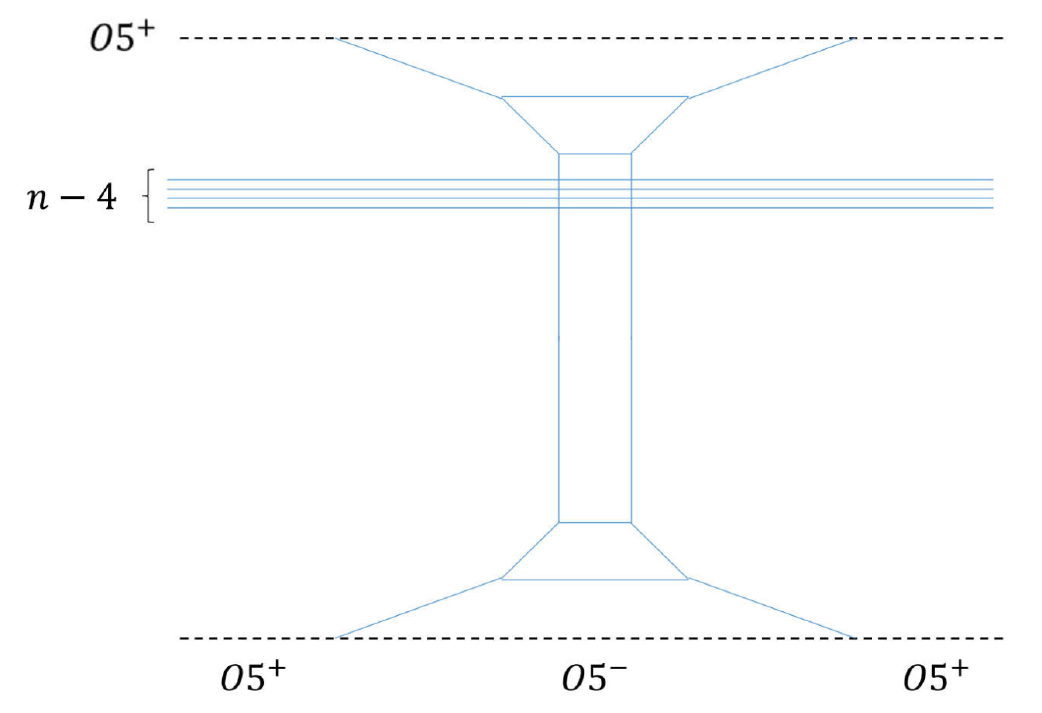} 
\caption{The brane web describing the effective 5d theory resulting from the 6d SCFT reduced on a circle.}
\label{SOdrWeb}
\end{figure}

We can consider compactifying one direction of the 6d SCFT, and taking the T-dual. Since the direction is in the 6d spacetime, it is shared by the D6, O$6$ and NS5-branes. As such, after the T-duality the D6-branes become D5-branes, the O$6^-$-plane becomes two O$5^-$-planes, and the two NS5-branes remain NS5-branes. The result is the brane web depicted in figure \ref{SOdrWeb}. We note that taking the radius of the circle to zero would correspond in the T-dual picture to taking the distance between the two O$5^-$-planes to infinity. As we wish to preserve the $usp(2N-16)$ global symmetry, this must be done while keeping the $N-8$ D5-branes on top of the same O$5$-plane (here we have chosen the upper one in the figure). We can use the brane web to read of the resulting theory. In this limit, we get two 5d SCFTs, corresponding to the upper and lower webs in figure \ref{SOdrWeb}, connected by gluing the two NS5-branes on top. The lower web in figure \ref{SOdrWeb} describes the 5d SCFT UV completing an $so(4)$ gauge theory, that is two decoupled copies of the $E_1$ SCFT. The upper web in figure \ref{SOdrWeb} describes the 5d SCFT UV completing an $so(N-4)+(N-8)V$ gauge theory. We shall say more about its properties latter, but for now we shall mention that it has an $su(2)\times usp(2N-16)$ flavor symmetry. Gluing the SCFTs by connecting the two NS5-branes corresponds to gauging an $su(2)$ flavor symmetry of the 5d SCFTs\footnote{This can be understood by using S-duality to map it to connecting parallel D5-branes.}.         

\begin{figure}
\center
\includegraphics[width=0.5\textwidth]{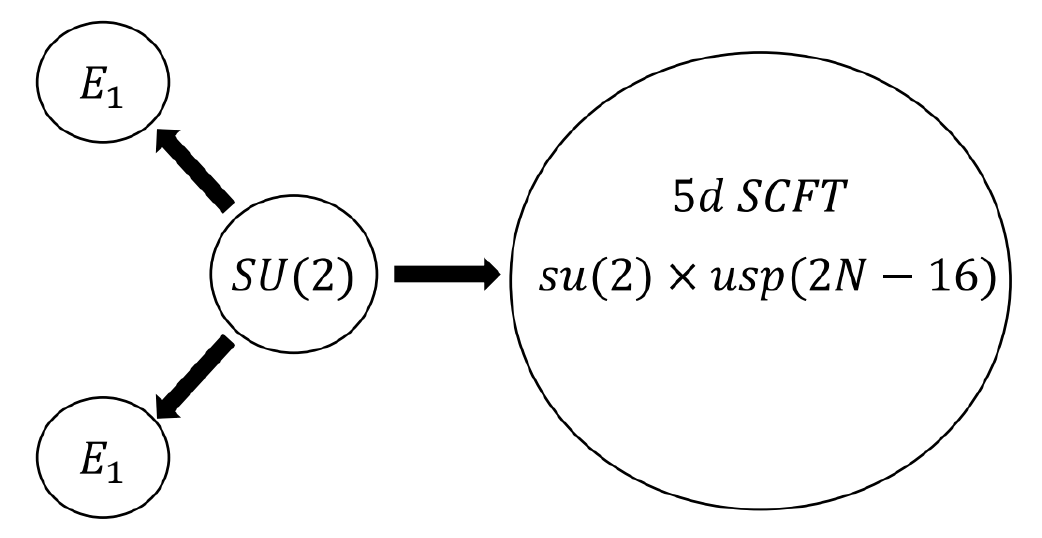} 
\caption{The 5d effective description of the 6d SCFT reduced on a circle. The central circle denotes an $SU(2)$ gauge theory, and the two smaller circles on the left denote $E_1$ SCFTs. The larger circle on the right denotes a 5d SCFT with $su(2)\times usp(2N-16)$ global symmetry. This 5d SCFT is the UV completion of an $so(N-4)+(N-8)V$ 5d gauge theory. The arrow denotes a gauging of the $SU(2)$ flavor symmetry of the 5d SCFTs by the $SU(2)$ gauge group (for the SCFT on the right the gauged symmetry is the $su(2)$ part of the flavor symmetry).}
\label{SONdr}
\end{figure}

 Overall, we arrive at the conclusion that the 6d SCFT reduced on a circle has an effective description as a 5d $SU(2)$ gauge theory gauging the $SO(3)$ global symmetry of $2$ disconnected $E_1$ 5d SCFTs and an additional 5d SCFT which is the UV completion of a $Spin(N-4)+(N-8)V$ gauge theory (gauging the $su(2)$ part of its global symmetry so that the $usp(2N-16)$ part is preserves), see figure \ref{SONdr}. Here we have chosen a specific global form that will match the global form and polarization chosen for the 6d SCFT on the circle. Before going into details about the matching of symmetries between the 5d theory and 6d SCFT, we should say a few words about the properties of the 5d $su(2)\times usp(2N-16)$ SCFT. These are based on the results of \cite{Apruzzi:2021vcu} and the 5d gauge theory description:

\begin{itemize} 
  \item The 5d SCFT has a continuous flavor symmetry forming the algebra $su(2)\times usp(2N-16)$. At the group level, the symmetry is $SO(3)\times USp(2N-16)$ for $N$ odd, and $SO(3)\times PUSp(2N-16)$ for $N$ even. In the above 5d theory, the $SO(3)$ is gauged by an $SU(2)$ gauge group.
	\item For $N$ even, the 5d gauge theory possesses a $\mathbb{Z}_2$ discrete symmetry associated with the outer automorphism of $Spin(N-4)$. We expect this symmetry to also be present in the 5d SCFT. We shall denote the background connection for it by $A_P$.
	\item The 5d SCFT has a $\mathbb{Z}_2$ 1-form symmetry which manifests in the gauge theory as the 1-form symmetry associated with the center of $Spin(N)$ that acts trivially on the vector . We shall denote the background connection for it by $B$. This 1-form symmetry is involved in a 2-group with the $SO(3)$ global symmetry, and the $PUSp(2N-16)$ one for $N=4n+2$. Additionally, for $N$ even, we also expect a 2-group involving $A_P \cup w_2(PUSp(2N-16))$. This is expected from the gauge theory description, and is needed to match the properties of the 5d and 6d theories. Overall, we expect a 2-group structure of $\delta B = \beta w_2(SO(3))$ for $N$ odd and $\delta B = \beta w_2(SO(3)) + n\beta w_2(PUSp(2N-16)) + A_P \cup w_2(PUSp(2N-16))$ for $N=2n$.
\end{itemize} 

Now we can properly study the mapping between the global symmetries of the 5d theory and 6d SCFT. We start with the continuous symmetries. First, the 5d theory has the supersymmetry algebra (the theory is not conformal) and $su(2)$ R-symmetry, mapping to the appropriate symmetries in 6d (conformal symmetry is broken due to the compactification). Additionally, there is the $u(1)$ instantonic symmetry of the $SU(2)$ gauge group which is associated with the KK modes on the circle. Finally, there is the $usp(2N-16)$ global symmetry coming from the 5d SCFT. This symmetry is globally $USp(2N-16)$ for $N$ odd and $PUSp(2N-16)$ for $N$ even, as expected from the 6d SCFT.

We next move to consider the discrete symmetries. First there is a $\mathbb{Z}_2$ symmetry associated with permuting the two $E_1$ SCFTs, and a $\mathbb{Z}_2$ symmetry of the 5d SCFT (for $N$ even) associated with the outer automorphism symmetry of its gauge theory description. Second there are $4$ $\mathbb{Z}_2$ 1-form symmetries, $2$ coming from the 1-form symmetries of the $E_1$ SCFTs, one from the center of the $SU(2)$ and one from the 5d SCFT\footnote{As before, we take the theta angle of the $SU(2)$ to be such that its instanton operators are uncharged under its center.}. From the 6d SCFT, we expect a $\mathbb{Z}_2$ 1-form symmetry coming from the 1-form symmetry of the 6d SCFT as well as a $\mathbb{Z}_4$ 1-form symmetry coming from the defect group on the circle. Additionally, there should be a $\mathbb{Z}_2$ 0-form symmetry coming from the 1-form symmetry on the circle and an additional one for $N$ even associated with the outer automorphism symmetry of the 6d SCFT.

We next consider the mapping between the symmetries, beginning with the 0-form symmetries. Here it is natural to identify the $\mathbb{Z}_2$ symmetry permuting the two $E_1$ SCFTs of the 5d theory with the 6d 1-form symmetry on the circle, as both exist for all $N$. This leads us to identify the $\mathbb{Z}_2$ symmetry associated with the outer automorphism symmetry of the 6d SCFT with the analogous one associated with the 5d $su(2)\times usp(2N-16)$ SCFT. As we shall soon see this identification allows us to naturally account for the higher group structure involving this symmetry.   

Next we move to consider the 1-form symmetry. Here is is convenient to start with the 6d 1-form symmetry. As previously mentioned, for $N$ even it is involved in a 2-group with the 0-form symmetries, particularly $w_2(USp(2N-16)/\mathbb{Z}_2)$, which should then reduce to a similar 2-group in 5d. Therefore, it is natural to identify its 5d reduction with the 1-form symmetry of the 5d $su(2)\times usp(2N-16)$ SCFT, as said SCFT possesses a similar 2-group structure (this is essentially as both have a gauge theory description as a $Spin(N)$ gauge group with $N$ differing between the two by $4$). There is also the additional 2-group in the 5d SCFT involving the $SO(3)$ that would be important next when we consider the $\mathbb{Z}_4$.

This leaves us with the $\mathbb{Z}_4$ 1-form symmetry expected from the 6d defect group. The story here is similar to the previous case. Specifically, both the $E_1$ SCFT and the $su(2)\times usp(2N-16)$ SCFT have a 2-group structure mixing $w_2(SO(3))$ with their 1-form symmetry. Here as the $SO(3)$ is gauged by an $SU(2)$, $w_2(SO(3))$ becomes identified with the background field for the electric 1-form associated with the $SU(2)$. The 2-group structure now leads to an extension of this $\mathbb{Z}_2$ 1-form symmetry and the diagonal 1-form symmetry of the two $E_1$ and $su(2)\times usp(2N-16)$ SCFTs to a $\mathbb{Z}_4$ 1-form symmetry. This is similar to the $Spin(8)$ case, though it is slightly simpler as there are fewer 1-form symmetries involved. As before, we are forced to identify the remaining 1-form symmetry as an accidental enhancement. 

Like in the previous case, the 3-group structure of the 6d SCFT should reduce to a 2-group structure of the 5d theory. For the case at hand it takes the form of: $\delta X_C = 2 A_{5d} \cup B_{5d}$, where we use $X_C$, $B_{5d}$ and $A_{5d}$ for the background fields for the 5d $\mathbb{Z}_4$ 1-form symmetry, $\mathbb{Z}_2$ 1-form symmetry and $\mathbb{Z}_2$ 0-form symmetry (coming from the 6d 1-form symmetry wrapping the circle), respectively. The 6d perspective then predicts the existence of this 2-group in the 5d theory. It would be interesting to study this explicitly in the 5d theory, though as before, we shall reserve this for future work.

Overall, we see that we can again map the symmetries between the 6d and 5d theories, though we are again forced to identify some symmetries as accidental. As before, we can try to study the anomalies of the 5d theory, though we shall not do so here. Finally, we can consider the reduction to 4d. This is tricky as it requires an understanding of what is the 4d reduction of the 5d $su(2)\times usp(2N-16)$ SCFT. We shall reserve this for future study, and instead consider the reduction of the 6d SCFT with a non-trivial Stiefel-Whitney class.  

\subsection{Stiefel-Whitney reduction}

We are now ready to consider the reduction on the torus with a non-trivial Stiefel-Whitney (SW) class, that is when $\int w_2(PUSp(2N-16))\neq 0$. As previously stated, this is especially interesting as $w_2(PUSp(2N-16))$ has non-trivial interplay with the rest of the symmetries, notably through higher group structures. Note that throughout this section $N$ is taken to be even, so the global symmetry is $PUSp(2N-16)$ and such a reduction makes sense.

We can tackle this problem as before by reducing first to 5d and then reduce the resulting 5d theory to 4d. To analyze this, we shall use a representation of the SW class as two almost commuting holonomies. The idea is to introduce the following two holonomies on the two independent cycles of the torus:

\bea \label{SWmatrices}
H_1 = i\begin{pmatrix}
  I & 0 \\
  0 & -I
\end{pmatrix} \; , \; H_2 = i\begin{pmatrix}
  0 & J \\
  J & 0
\end{pmatrix} .
\eea 

Note that both matrices are in $USp(4n-16)$ and they obey: $H^{-1}_1 H^{-1}_2 H_1 H_2 = -I$, that is the holonomies commute up to the central element of $USp(4n-16)$. Since the homotopy group relation of the torus enforces that the holonomies around the two basic cycles must commute, these holonomies define an acceptable $PUSp(4n-16)$ background that cannot be lifted to $USp(4n-16)$.

The presence of the holonomies leads to the breaking of the flavor symmetry, in this case to $USp(2n-8)$ \footnote{This is the maximal possible symmetry one can preserve, though other choices exist. For instance, it is possible to preserve $SO(2n-8)$ instead, or $USp(2n-2m-8)\times SO(2m)$ more generally.}. It is instructive to see this from the explicit expressions. First let us consider $H_1$. The general matrices that commute with it have the form:

\be
M=\begin{pmatrix}
  M_1 & 0 \\
  0 & M_2
\end{pmatrix} .
\ee   

The demand that $M$ be unitary forces $M_1$ and $M_2$ to be unitary, and the demand that $M$ be in $USp(4n-16)$ further enforces that:

\bea
&& M^TJM=J \rightarrow \begin{pmatrix}
  M^T_1 & 0 \\
  0 & M^T_2
\end{pmatrix} \begin{pmatrix}
  0 & I \\
  -I & 0
\end{pmatrix} \begin{pmatrix}
  M_1 & 0 \\
  0 & M_2
\end{pmatrix} = \begin{pmatrix}
  0 & I \\
  -I & 0
\end{pmatrix} \rightarrow \nonumber \\ && \begin{pmatrix}
  0 & M^T_1 M_2 \\
  -M^T_2 M_1 & 0
\end{pmatrix} = \begin{pmatrix}
  0 & I \\
  -I & 0
\end{pmatrix} .
\eea   

This suggests that $M_2 = M^*_1$. As such, we see that the commutant of $H_1$ in $USp(4n-16)$ is $U(2n-8)$. However, that is not the full symmetry that is preserved. Note that $H_2$ acts on $M$ by $M_1 \leftrightarrow -J M_2 J$, that is $M_1 \rightarrow -J M^*_1 J$. As such, while it does not commute with $M$, it preserve its form. In mathematical terms, we say that $H_2$ is part of the normalizer of $H_1$ in $USp(4n-16)$ though it is not in the commutant (centralizer). As the preserved symmetry is generally the normalizer, we see that $H_1$ breaks $USp(4n-16)$ to $U(2n-8) \rtimes \mathbb{Z}_2$, where the $\mathbb{Z}_2$ acts on $U(2n-8)$ via its charge conjugation outer automorphism\footnote{Note that the action $M_1 \rightarrow -J M^*_1 J$ is equivalent to charge conjugating $M_1$, which is an outer automorphism, followed by acting on it with the uniary matrix $J$ which is an inner automorphism and so is included in $U(2n-8)$.}.

Now if we also consider $H_2$, we see that we must further restrict to those $M$ for which $M_1 = -J M^*_1 J \rightarrow J M_1 = M^*_1 J$, which further breaks $U(2n-8)$ to $USp(2n-8)$ \footnote{It is possible to get other real subgroups of $U(2n-8)$ by a different choice of $H_2$. For instance, if we were not to conjugate by $J$ then the preserved group would be $SO(2n-8)$, with more general choices presering $USp(2n-2m-8)\times SO(2m)$, see \cite{Tachikawa:2011ch}. Here we have specialized to the representation of $w_2$ preserving the $USp(2n-8)$ subgroup.}.

We can now consider the reduction to 4d on a torus with a non-trivial SW class. We can represent it by the two almost commuting holonomies, $H_1$ and $H_2$, and analyze the reduction by first reducing on a circle to 5d, and then continuing to 4d. 

\subsubsection{Reduction to 5d}

We begin with the circle reduction to 5d. Here we essentially reduce on a circle with the holonomy $H_1$, which implements the breaking of $usp(4n-16)\rightarrow u(2n-8)\rtimes \mathbb{Z}_2$. The holonomy is then interpreted as a mass deformation of the 5d theory found above, implementing such breaking. We can use the brane realization of the 5d theory to find the 5d theory after the deformation. Specifically, the deformation can be described by pulling the D5-branes away from the O$5$-planes, which breaks the global symmetry on them to $u(2n-8)$. This suggests that the deformation amounts to pulling the D5-branes so that they sit exactly in the middle between the two orientifolds, as that is the furthest we can bring them from the orientifolds. Furthermore, at that point we get a $\mathbb{Z}_2$ symmetry associated with reflecting the diagram. As we shall soon argue, this symmetry is the charge conjugation symmetry inside $usp(4n-16)$.

\begin{figure}
\center
\includegraphics[width=0.5\textwidth]{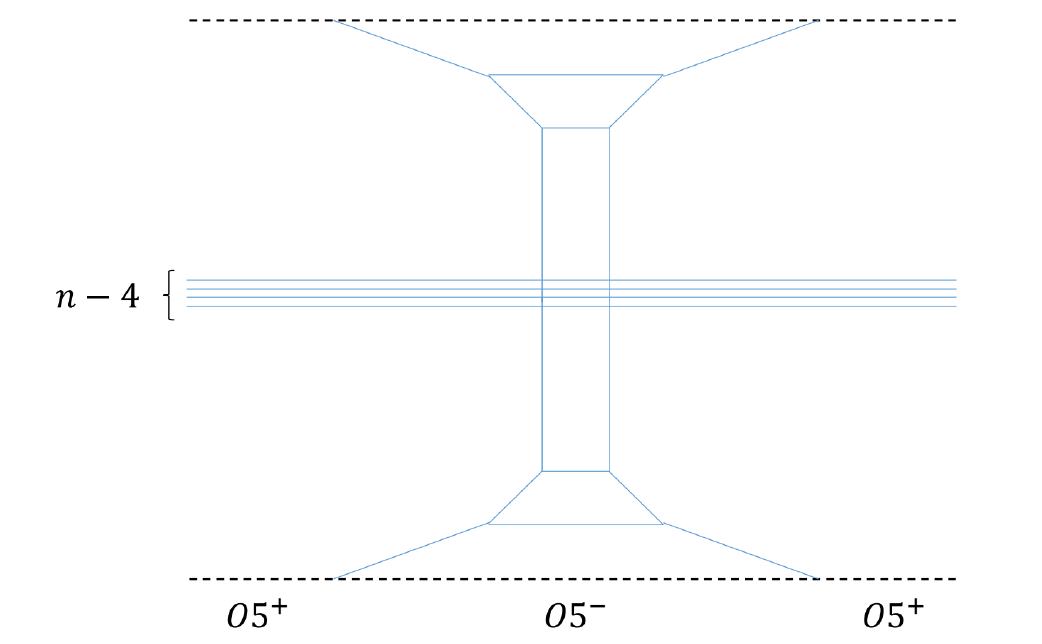} 
\caption{The brane web describing the effective 5d theory resulting from the 6d SCFT reduced on a circle.}
\label{SOswWeb}
\end{figure}

The resulting web diagram is depicted in figure \ref{SOswWeb}. Once we take the distance between the orientifolds to be very large, corresponding to taking the compactification radius to be very small, the brane web can be broken to the lower, middle, and upper diagrams, which are connected to one another by two long NS5-branes. As we previously mentioned the long NS5-branes are associated with a 5d $SU(2)$ gauge theory, and the gluing implies that these gauge an $su(2)$ global symmetry of the 5d SCFTs they are attached to. Here we have two different types of 5d SCFTs. The lower and upper brane diagrams in figure \ref{SOswWeb} describe two decoupled $E_1$ SCFTs, as before. The middle brane diagrams in figure \ref{SOswWeb} describes an SCFT associated with the intersection of two NS5-branes and $n-4$ D5-branes. This describes a 5d SCFT with $su(2)\times su(2)\times su(2n-8)$ global symmetry. This SCFT can be described as the UV completion of the 5d gauge theory $SU(n-4)_0+(2n-8)F$ (here the subscript denotes the Chern-Simons level).

\begin{figure}
\center
\includegraphics[width=0.75\textwidth]{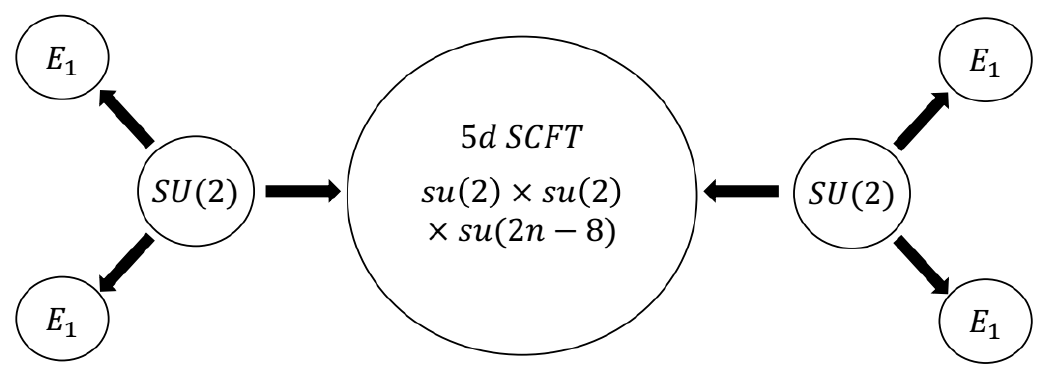} 
\caption{The 5d effective description of the 6d SCFT reduced on a circle with the holonomy $H_1$. The central large circle is a 5d SCFT with $su(2)\times su(2)\times su(2n-8)$ global symmetry. This 5d SCFT is the UV completion of an $SU(n-4)_0+(2n-8)F$ 5d gauge theory. The remaining circles denote $SU(2)$ gauge groups and $E_1$ SCFTs, as before. Also as before, the arrows denote gauging of an $SU(2)$ factor of the flavor symmetry by the $SU(2)$ gauge group.}
\label{SONsw}
\end{figure}

Overall, we see that the resulting 5d theory can be described by two $SU(2)$ gauge groups, each gauging two $E_1$ SCFTs, and both are connected by gauging the $su(2)$ global symmetry of the 5d SCFT with $su(2)\times su(2)\times su(2n-8)$ global symmetry. An illustration of the resulting theory is given in figure \ref{SONsw}. Next we wish to state a few properties of the $su(2)\times su(2)\times su(2n-8)$ SCFT that would be useful later:

\begin{itemize} 
  \item The 5d SCFT has a continuous flavor symmetry forming the algebra $su(2)\times su(2)\times su(2n-8)$. It can be realized as the UV completion of at least two different 5d gauge theories. One is the previously mentioned $SU(n-4)_0+(2n-8)F$, where the $su(2n-8)$ part of the global symmetry is manifest, but only the Cartan of the $su(2)\times su(2)$ part. The other 5d gauge theory is a long quiver of $n-5$ $SU(2)$ gauge groups, connected by bifundamentals, with two flavors for each of the edge groups. Here the $su(2)\times su(2)$ part is manifest but not the $su(2n-8)$ part, where only its $su(2)^{n-4}\times u(1)^{n-5}$ subgroup is manifest. 
	
	The above gauge theories can be used to infer various properties of the 5d SCFT. Specifically, they suggest that at the group level the $su(2)\times su(2)$ part of the flavor symmetry is actually $[SU(2)\times SU(2)]/\mathbb{Z}_2$ and that the $su(2n-8)$ part is globally $SU(2n-8)/\mathbb{Z}_{n-4}$. One can study the spectrum of Higgs branch operators, for instance using the magnetic quiver in \cite{Ferlito:2017xdq}, which support this identification, and suggests a further diagonal $\mathbb{Z}_2$ quotient between $[SU(2)\times SU(2)]/\mathbb{Z}_2$ and $SU(2n-8)/\mathbb{Z}_{n-4}$. Specifically, there is a Higgs branch chiral ring operator in the $(\bf{ 2}, \bf{ 2} , \bf{ \Lambda}^{n-4})$ of the $su(2)\times su(2)\times su(2n-8)$ global symmetry, which is consistent with the above identification.	
	\item The 5d SCFT does not possess any 1-form symmetries, as is apparent from the gauge theory descriptions. However, it possess a discrete $\mathbb{Z}_2$ symmetry. This symmetry can be seen from the web as a rotation of the web diagram, see \cite{Zafrir:2016wkk}. In the gauge theory descriptions, it appears as the charge conjugation symmetry of the $SU(n-4)_0+(2n-8)F$ theory or the quiver reflection symmetry of the $SU(2)$ quiver. This suggests that this symmetry acts on the global symmetry by exchanging the two $su(2)$ groups and charge conjugating the $su(2n-8)$ part. 
\end{itemize}

We next study the global symmetry of the 5d theory, and compare against the expectations from the 6d description. The continuous global symmetry includes (in addition to the 5d supersymmetry algebra and $SU(2)$ R-symmetry) the $su(2n-8)$ global symmetry, and two $U(1)$ groups coming from the topological symmetries of the two $SU(2)$ gauge groups. These should map to the $u(2n-8)$ that is the commutant of $H_1$ in $usp(4n-16)$ and the $U(1)$ associated with the KK modes. As we shall argue soon, the precise mapping should be that the diagonal combination of the two instanton $U(1)$ groups should map to the $U(1)$ associated with the KK modes, while the anti-diagonal combination should map the $U(1)$ commutant in $usp(4n-16)$.

Next we move to discuss the discrete symmetries. The model has several interesting discrete symmetries. First, there is the symmetries exchanging the two $E_1$ SCFTs on the two sides. These give a $\mathbb{Z}_2 \times \mathbb{Z}_2$ 0-form symmetry. Additionally, there is a $\mathbb{Z}_2$ symmetry associated with reflecting the quiver. Here, it is important that the $su(2)\times su(2)\times su(2n-8)$ 5d SCFT has a $\mathbb{Z}_2$ symmetry exchanging the two $su(2)$ groups and charge conjugating the $su(2n-8)$. Due to the charge conjugation, it is natural to identify this symmetry with the $\mathbb{Z}_2$ in $usp(4n-16)$ (that is in the normalizer but not the centralizer)\footnote{Recall that this symmetry should also charge conjugate the $U(1)$ commutant of $H_1$ in $usp(4n-16)$. This is why it was natural to identify it with the anti-diagonal combination of the instantonic $U(1)$ groups.}. The remaining two $\mathbb{Z}_2$ groups are then identified with the outer automorphism symmetry of the 6d SCFT and the 1-form symmetry wrapping the circle.

However, note that the symmetries in 5d actually combine to form a larger group. Specifically, the symmetry reflecting the entire 5d theory also exchanges the two $\mathbb{Z}_2$ symmetries acting by the exchange the two $E_1$ SCFTs. As such, all three symmetries should combine to form $(\mathbb{Z}_2 \times \mathbb{Z}_2) \rtimes \mathbb{Z}_2 = D_4$. So in fact we have a dihedral group in 5d, but from where does it come from in 6d? We shall next address this issue.

We can argue for this extension as follows. We begin by noting that due to the holonomy, $w_2$ wrapping the circle reduces to the background gauge field for the charge conjugation symmetry. Recall that in the presence of $H_1$ and $H_2$ on the torus we have that $\int_{T^2} w_2 = 1$. As such, the integral of $w_2$ on any one of the circles cannot be zero. Also recall that following the breaking of the global symmetry induced by the two holonomies, $H_2$ can be interpreted as an holonomy for the charge conjugation symmetry. This suggests that in its presence: $\int_{S^1_{H_2}} A_C = 1$, with $A_C$ the background gauge field for the charge conjugation symmetry. Consistency now requires: $\int_{S^1_{H_1}} w_2 = A_C$. As before, we also have that the 1-form symmetry connection, $B$, reduces to the connection for the 0-form symmetry when wrapping the circle: $\int_{S^1} B = A_B$.   

We then see that both $B$ and $w_2$ wrapping the circle reduce to background gauge fields for 0-form symmetries. Recall that the two symmetries, together with the outer automorphism symmetry, are involved in a 2-group: $\delta B = n\beta w_2 + A_P \cup w_2$. However, as the integral of $w_2$ on the circle now does not vanish, this 2-group reduces to a non-trivial extension of the 0-form symmetries. As before, it is convenient to analyze the reduction using a mock symmetry TQFT. One complication here is how to deal with $w_2$. Here we are only interested in determining the reduction of the 2-group relation, which we expect would be the same should $w_2$ be just a standard background connection for a $\mathbb{Z}_2$ 1-form symmetry. As such, we shall merely treat it as such for the purpose of this computation. We then write the following action for the mock symmetry TQFT:   

\be \label{6dSymTFT}
\pi i \int c_4 \cup \delta A_P + c \cup \delta B + w_2 \cup \delta c_w + n c \cup \beta w_2 + c \cup A_P \cup w_2 ,
\ee
where $c_4$, $c$ and $c_w$ are the dual gauge field to $A_P$, $B$ and $w_2$, respectively. 

We can next reduce the symmetry TFT on the circle where we get:

\bea 
&& \pi i \int c_3 \cup \delta A_P + b \cup \delta B + c \cup \delta A_B + A_C \cup \delta c_w + w_2 \cup \delta b_w  + n b \cup \beta w_2 \nonumber \\ &&  + n c \cup A_C \cup A_C + b \cup A_P \cup w_2 + c \cup A_P \cup A_C , \label{5dSymTFT}
\eea
where here we used:

\bea
\int_{M_7} c \cup \beta w_2 & = & \int_{M_7} c \cup Sq^1 (w_2) = \int_{M_7} Sq^1(c \cup w_2) + Sq^1(c) \cup w_2 \\ \nonumber & = & \int_{M_6} Sq^1(c) \cup A_C = \int_{M_6} Sq^1(c \cup A_C) + c \cup Sq^1 (A_C) = \int_{M_6} c \cup A_C \cup A_C . 
\eea
and we have assumed that all the manifolds are spin (so $Sq^1(c \cup w_2) = w'_1\cup c \cup w_2 =0$).  

We can rewrite \eqref{5dSymTFT} as:

\bea
&& \pi i \int c_3 \cup \delta A_P + b \cup (\delta B + n\beta w_2 + A_P \cup w_2) + c \cup (\delta A_B + n A_C \cup A_C + \cup A_P \cup A_C) \nonumber \\ && + A_C \cup \delta c_w + w_2 \cup \delta b_w .
\eea 

From this we see that the 6d 2-group leads to two interesting effects in the 5d theory. The first is a similar 2-group in 5d involving the 1-form, (whatever is left of) $w_2$ and the 5d reduction of the 6d outer automorphism symmetry: $\delta B = n\beta w_2 + A_P \cup w_2$. The second is a standard central extension involving the three 5d $\mathbb{Z}_2$ 0-form symmetries: $\delta A_B = n A_C \cup A_C + A_P \cup A_C$. This extension is known to give the dihedral group $D_4$ regardless of whether $n$ is zero or one \cite{Bergman:2024its} (see also appendix \ref{App:D4}), so we see that the reduction of the 6d symmetry structure should indeed give the discrete group $D_4$ in 5d, as observed. The structure of the extension also suggests that the central element, here the simultaneous exchange of both pairs of $E_1$ SCFTs, should be identified with the 1-form symmetry on the circle. 

This brings us to the 1-form symmetries. The 5d model has $5$ $\mathbb{Z}_2$ 1-form symmetries, $4$ coming from the $E_1$ SCFTs, and one coming from the $SU(2)$ gauge groups. Note that here we have two $SU(2)$ groups which are connected via the 5d SCFT. Since the $su(2)\times su(2)$ part of the flavor symmetry of the 5d SCFT should be $[SU(2)\times SU(2)]/\mathbb{Z}_2$, the 5d SCFT provides matter charged under the $\mathbb{Z}_2$ center of each individual $SU(2)$ gauge group, but not under the diagonal one. This suggests the presence of a $\mathbb{Z}_2$ 1-form symmetry coming from the two $SU(2)$ groups\footnote{As before, we assume the $SU(2)$ theta angles are such that the 1-form symmetry is not broken by their instanton operators.}.  

 As before, due to the 2-group structure of the $E_1$ SCFTs, two of the $\mathbb{Z}_2$ 1-form symmetries combine to form a $\mathbb{Z}_4$, which is the reduction of the 6d defect group. This should be the electric 1-form symmetry associated with the $SU(2)$ groups and the diagonal combination of the 1-form symmetries associated with all four $E_1$ SCFTs. This follows similarly to the discussion for the $Spin(8)$ case so we won't repeat the analysis here. Note that this 1-form symmetry commutes with the dihedral 0-form symmetry, as expected. This leaves us with three $\mathbb{Z}_2$ 1-form symmetries. One combination of these should be the reduction of the 6d 1-form symmetry, while the remaining two appear to be accidental enhancements. 

Next, we turn to the fate of the 2-group for the 1-form symmetries. The previous analysis of the mock symmetry TQFT, \eqref{5dSymTFT}, suggests that in addition to the extension involving the 0-form symmetries, the 2-group should also reduce to a similar one involving the 1-form symmetries: $\delta B = n\beta w^{(5d)}_2 + A_P \cup w^{(5d)}_2$, where $w^{(5d)}_2$ is the reduction of the 6d SW class. One subtlety here is that the global symmetry is broken due to $H_1$, so we need to carefully consider what happens to $w_2$.  

Recall that $H_1$ breaks $usp(4n-16)$ to $u(2n-8)$ such that: $({\bf 4n-16})_{usp(4n-16)} \rightarrow q ({\bf 2n-8})_{su(2n-8)} + \frac{1}{q} (\overline{\bf 2n-8})_{su(2n-8)}$. This suggests that $w_2$ can be embedded in either $C_1(U(1))$ mod 2, or in $w_2 (SU(2n-8)/\mathbb{Z}_2)$. We would then expect that $w_2 (PUSp(4n-16)) \rightarrow \tilde{C}_1(U(1)) + w_2 (SU(2n-8)/\mathbb{Z}_2)$. This appears to suggests that in 5d we should have the 2-group $\delta  B^{(5d)} = n\beta w_2 (SU(2n-8)/\mathbb{Z}_2) + A_P \cup w_2 (SU(2n-8)/\mathbb{Z}_2) + A_P \cup \tilde{C}_1(U(1))$, where we used the fact that $\beta \tilde{C}_1(U(1)) = 0$. We also note that the decomposition of the fundamental of $usp(4n-16)$ to $u(2n-8)$ suggests that at the group level the breaking is $USp(4n-16) \rightarrow [U(1)\times SU(2n-8)]/\mathbb{Z}_{2n-8}$.

In the 5d theory, the $su(2n-8)$ part comes from the 5d SCFT, while the $U(1)$ should be the anti-diagonal combination of the instantonic $U(1)$ symmetries of the two $SU(2)$ gauge groups\footnote{Recall that the charge conjugation symmetry, $A_C$, should have a non-trivial action on this $U(1)$. As this symmetry is expected to map to the quiver reflection symmetry, the $U(1)$ should be the anti-diagonal combination so that we get the right action.}. Naively, at the group level the $su(2n-8)$ part should be $SU(2n-8)/\mathbb{Z}_{2n-8}$, as the matter spectrum in the 5d SCFT seems consistent with this, once the $SU(2)\times SU(2)$ part is gauged. However, due to said gauging, the 5d SCFT provides matter for the two $SU(2)$ groups. Thus, the instantons of the two $SU(2)$ groups can acquire charges under the $su(2n-8)$ part. This should presumably lead to the flavor symmetry being the diagonal quotient between the two, though this is beyond our ability to compute at this moment. At low-energies, where the instanton particles are integrated out, the symmetry appears to reduce to the $SU(2n-8)/\mathbb{Z}_{2n-8}$ that we observe.  

Next, we shall argue that the 5d low-energy theory indeed possess the 2-group $\delta B = \beta w_2(SU(2n-8)/\mathbb{Z}_{2})$ when $n$ is odd. The idea is to consider coupling the 5d theory to a non-trivial $w_2(SU(2n-8)/\mathbb{Z}_{2})$. As we have mentioned the global symmetry of the 5d SCFT contains $[SU(2)\times SU(2)]/\mathbb{Z}_2$ and $SU(2n-8)/\mathbb{Z}_{n-4}$ with an additional diagonal $\mathbb{Z}_{2}$ quotient. Because of said diagonal quotient, the global symmetry becomes $SU(2n-8)/\mathbb{Z}_{2n-8}$ once the $SU(2)\times SU(2)$ part is gauged. When $n$ is even, $w_2(SU(2n-8)/\mathbb{Z}_{2})$ can be completely embedded as a non-trivial $SU(2n-8)/\mathbb{Z}_{n-4}$ bundle in the 5d SCFT. On the the hand for $n$ odd, $w_2(SU(2n-8)/\mathbb{Z}_{2})$ is embedded in the diagonal quotient. As such, its presence also necessitates the presence of a non-trivial $w_2(SO(3))$ for one of the two $SU(2)$ gauge groups, that is we are forced to set $w_2(SO(3)) = w_2(SU(2n-8)/\mathbb{Z}_{2})$. However, the $SU(2)$ also gauges two $E_1$ SCFTs which identifies $w_2(SO(3))$ of their flavor symmetry with that of the $SU(2)$ gauge symmetry. Recall that the $E_1$ SCFTs possess the 2-group: $\delta B_i = \beta w_2(SO(3))$. The above identification then leads to the 2-group $\delta B_i = \beta w_2(SU(2n-8)/\mathbb{Z}_{2})$ for the two adjacent $E_1$ SCFTs. Overall, we see that in the $n$ odd case we get the 2-group $\delta B = \beta w_2(SU(2n-8)/\mathbb{Z}_{2})$, where here $B$ is the background field for the diagonal 1-form symmetry of two adjacent $E_1$ SCFTs.

We see then how we can recover at least part of the expected 2-group structure in the 5d theory. Nevertheless, we should mention that there are a few subtle issues that were glanced over in the above analysis. First, the analysis was done in the low-energy 5d theory, so we ignored instantons of the $SU(2)$ gauge groups, though these should be charged under the $su(2n-8)$ symmetry, and so could affect the analysis. Yet, this seems reasonable as we expect the 2-group structure (at least the parts independent of the $U(1)$) to be preserved in the RG flow, so we might as well look for it at low-energy. Second, as the $su(2n-8)$ symmetry is inside a 5d SCFT, there might be additional effects when coupling it to $w_2$. For instance, consider deforming the 5d SCFT so that it flows to the $SU(n-4)_0+(2n-8)F$ gauge theory. Note that the matter in the gauge theory is in the fundamental of the $su(2n-8)$ flavor symmetry, so coupling to $w_2(SU(2n-8)/\mathbb{Z}_{2})$ must be accompanied by a non-trivial $w_2(SU(n-4)/\mathbb{Z}_{2})$ for the gauge symmetry. In the presence of the latter, fractional instantons are allowed so the 5d SCFT may exhibit additional excitations in its presence, which may affect the analysis.

Our point here is that we can recover the 2-group $\delta B = n\beta w_2(SU(2n-8)/\mathbb{Z}_{2})$ from this simpler analysis, though a more complete one would require also considering these more subtle issues. While it will be interesting to perform such an analysis, we shall reserve it to future work. Additionally,  as previously said, the 6d analysis suggests a larger 2-group of the form $\delta  B^{(5d)} = n\beta w_2 (SU(2n-8)/\mathbb{Z}_2) + A_P \cup w_2 (SU(2n-8)/\mathbb{Z}_2) + A_P \cup \tilde{C}_1(U(1))$. It would be interesting to argue for this explicitly from the 5d theory, though we shall reserve this as well for future work. 

Finally, there is also the 6d 3-group structure, which as before should lead to the 5d 2-group structure $\delta X_C = 2 A_B \cup B$, and it would be interesting to verify its existence directly in 5d. We again defer this to future work, though we shall see some of the implications of this structure in the 4d reduction.  

Overall, we see that we can indeed match (most of) the global symmetry structure between the 6d and 5d theories.

\subsubsection{Reduction to 4d}

\begin{figure}
\center
\includegraphics[width=0.75\textwidth]{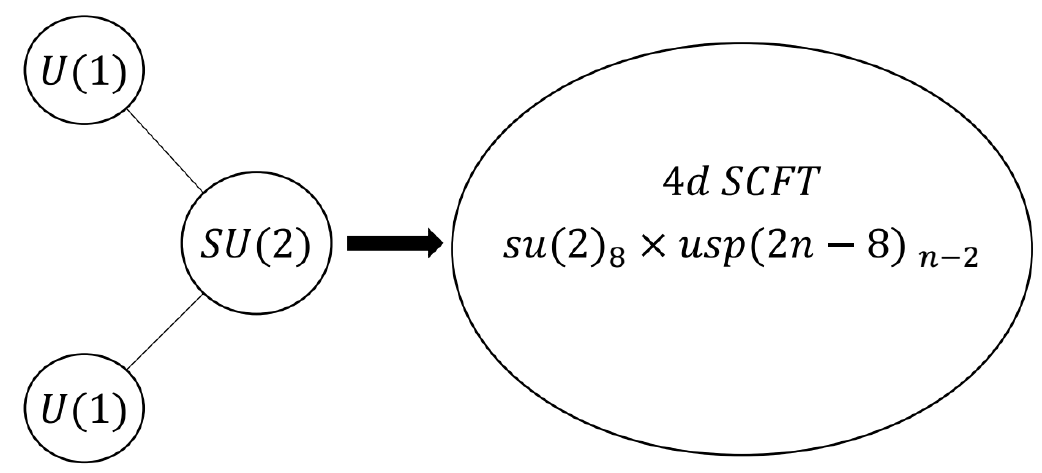} 
\caption{The 4d theory resulting from the reduction of the 6d SCFT on a torus with a non trivial Stiefel-Whitney class. The central large circle is a 4d SCFT with $su(2)\times usp(2n-8)$ global symmetry, and the subscript denote the flavor central charge. The remaining circles denote an $SU(2)$ gauge group and two $U(1)$ gauge groups. As before, the arrow denote gauging of an $SU(2)$ factor of the flavor symmetry by the $SU(2)$ gauge group, and lines between gauge group denote bifundamental hypermultiplets.}
\label{SONsw4d}
\end{figure}

We next consider the reduction on the circle to 4d with a twist by the 5d quiver reflection symmetry (we can also consider the non-twisted reduction though we shall not do so here). As noted this should give the 4d theory resulting from the torus compactification of the 6d SCFT with a non-trivial $w_2(PUSp(4n-16))$. The resulting 4d theory can be inferred from the effective 5d theory in figure \ref{SONsw}. Specifically, the twist by the quiver reflection symmetry identifies the two $SU(2)$ gauge groups and the two pairs of $E_1$ SCFTs. As such these reduce to a 4d $SU(2)$ gauge theory gauging the 4d twisted reduction of the 5d SCFT and the regular reduction of a pair of $E_1$ SCFTs. The latter we have seen reduce to an $\mathcal{N}=2$ $U(1)$ gauge theory with two hypermultiplets. As for the former, the twisted reduction of the 5d SCFT is believed to yield a 4d SCFT with $su(2)\times usp(2n-8)$ global symmetry \cite{Zafrir:2016wkk}. Here the $su(2)\times su(2)\times su(2n-8)$ is projected by the twist, leading to the $su(2)\times usp(2n-8)$ global symmetry of the 4d SCFT\footnote{As previously mentioned, the twist acts on the $su(2n-8)$ factor of the global symmetry group by its charge conjugation outer automorphism, projecting to a real subgroup. It is possible to change the preserved subgroup by conjugation with an $su(2n-8)$ element. Here we assume a specific choice is made such that the $usp(2n-8)$ subgroup is preserved. This matches the specific choice of the holonomies implementing the Stiefel-Whitney twist.}. Said 4d SCFT has the property that gauging the $su(2)$ part of the flavor symmetry is conformal and dual to the 4d $\mathcal{N}=2$ conformal gauge theory $so(n-2)+(n-4)V$ . Overall, the resulting 4d theory is illustrated in figure \ref{SONsw4d}.   

Next we shall examine the discrete global symmetries of the 4d theory. Specifically, we consider what happens to the symmetries of the 5d theory under the twisted reduction and attempt to interpret it in terms of the Stiefel-Whitney reduction of the 6d SCFT. We begin with the discrete 0-form symmetries. As we have seen previously, the 5d theory has a $D_4$ symmetry made from the 1-form symmetry on the circle, outer automorphism symmetry and quiver reflection symmetry that are extended due to the 6d 2-group structure. Here we twist by the quiver reflection symmetry, which would then act trivially in 4d. Additionally, symmetries that do not commute with it are broken. This suggests that only the 1-form symmetry on the circle, which is central in $D_4$, survives in 4d, while the outer automorphism symmetry is broken.

We can consider the same effect from the 6d viewpoint. Here the quiver reflection symmetry is part of the $PUSp(4n-16)$ symmetry, and is broken by the non-trivial Stiefel-Whitney class. However, what breaks the outer automorphism symmetry? To understand this, we recall that there is a 2-group relation in the 6d SCFT of the form: $\delta B = n\beta w_2 + A_P \cup w_2$. Note that due to the second term, this relation would not be invariant under background gauge transformations of $A_P \rightarrow A_P + \delta \lambda$, unless we also shift $B \rightarrow B + \lambda w_2$ \footnote{This can also be seen by looking at the gauge transformation of $A_P$ that leaves the mock symmetry TQFT, \eqref{6dSymTFT}, invariant}. This suggests that in the presence of a non trivial $w_2$, the holonomy of $B$ on the torus will not be invariant under background gauge transformations of $A_P$, as: $\int_{T^2} B \rightarrow \int_{T^2} B + \int_{T^2} \lambda w_2 = \int_{T^2} B + \lambda$. Since the holonomy of $B$ becomes a parameter in the 4d theory, this suggests that the 4d theory is not invariant under background gauge transformations of $A_P$, and as such that the outer automorphism symmetry should be broken in the reduction, as observed. We note that similar effects were observed for continuous symmetries in \cite{Nardoni:2024sos}. 

For $n$ odd there is also the 2-group part involving $\beta w_2$, and it is interesting if this term also leads to peculiar effects if a compactification is done with a non-trivial $w_2$. Here we note that the continuous global symmetry in the resulting 4d theory appears to be $PUSp(2n-8)$ for $n$ even but $USp(2n-8)$ for $n$ odd. This comes about as the $SU(2)$ gauging of the 4d $su(2)_8\times usp(2n-8)_{n-2}$ SCFT is dual to an $so(n-2)+(n-4)V$ gauge theory, and the global symmetry of the latter is $USp(2n-8)$ when $n$ is odd. We also note that here the 6d $usp(4n-16)$ symmetry is broken to $U(1)\times usp(2n-8)$ as $({\bf 4n-16}) \rightarrow (q + \frac{1}{q})({\bf 2n-8})$, suggesting that the 6d $PUSp(4n-16)$ symmetry is broken to $PUSp(2n-8)$. Yet, in the 4d theory we seem to observe $USp(2n-8)$ for $n$ odd, or in other words, the option of turning on a non-trivial $w_2$ is lost in this case. It is tempting to associate this with the 2-group involving $\beta w_2$ for $n$ odd. Specifically, we have seen that in a mixed 2-group between $w_2$ and $A_P$, a non-trivial $w_2$ leads to the breaking of $A_P$. We can think of the term $\beta w_2$ as a 2-group involving only $w_2$, which may suggests that in its presence a non-trivial $w_2$ on the surface leads to the "breaking" of $w_2$ in 4d. This is consistent with our observations in the 4d theory.

This leaves us with the  4d 0-form symmetries coming from the reduction of the 5d 1-form symmetries on the circle. As we have seen previously, the 5d 1-form symmetry associated with an $SU(2)$ gauge group reduces only to a 4d 1-form symmetry, but the 1-form symmetry of an $E_1$ SCFT also reduces to a 0-form symmetry, given by the charge conjugation of the 4d $U(1)$ gauge group. As such, we see that the 5d 1-form symmetry reduces to a $\mathbb{Z}_2 \times \mathbb{Z}_2$ 0-form symmetry associate with the individual charge conjugation of each $U(1)$ gauge group. This is indeed a symmetry of the 4d theory from essentially the same arguments as in section \ref{SO84dR}. Note that this $\mathbb{Z}_2 \times \mathbb{Z}_2$ 0-form symmetry does not commute with the $\mathbb{Z}_2$ symmetry exchanging the two $U(1)$ gauge groups, and all three combine to form the dihedral group $D_4$.

Next we want to explain these features from 6d. As we previously argued, the outer automorphism symmetry of the 6d SCFT is broken by the 2-group in the presence of the non-trivial Stiefel-Whitney class. This leaves us only with discrete 0-form symmetries coming from the higher form symmetries of the 6d SCFT. Specifically, we have a $\mathbb{Z}_2$ 1-form symmetry from which we expect to get a $\mathbb{Z}_2 \times \mathbb{Z}_2$ 0-form symmetry in 4d from wrapping the two cycles of the torus. Additionally, we have the $\mathbb{Z}_4$ self-dual 2-form symmetry, which when wrapped on the torus should give a 4d $\mathbb{Z}_4$ 0-form symmetry. However, here the $\mathbb{Z}_4$ self-dual 2-form symmetry gives only a  4d $\mathbb{Z}_2$ 0-form symmetry, associated with the subgroup. This follows as the 6d 2-form symmetry reduces to a 5d 1-form symmetry, made from the 1-form symmetry associated with the $SU(2)$ groups and the diagonal 1-form symmetry of the $E_1$ SCFTs, and the former does not reduce to a 0-form symmetry in 4d. The diagonal 1-form symmetry of the $E_1$ SCFTs should reduce to a 4d $\mathbb{Z}_2$ 0-form, which here is the simultaneous charge conjugation of the two $U(1)$ gauge group (or just the entire 4d theory, as the rest of the matter should be real). 

As such, we see that in 4d we get three $\mathbb{Z}_2$ 0-form symmetries, associated with the charge conjugation and exchange of the two $U(1)$ gauge groups. The simultaneous charge conjugation should map to the $\mathbb{Z}_2$ subgroup of the 6d $\mathbb{Z}_4$ self-dual 2-form symmetry wrapping the torus (with the $\mathbb{Z}_2$ quotient part being spontaneously broken). The symmetry exchanging the two $U(1)$ groups should map to the 6d 1-form symmetry wrapping the first cycle (this essentially follows from the 5d picture), which suggests that the charge conjugation of a single $U(1)$ should be the 6d 1-form symmetry wrapping the other cycle. However, we have seen that the three $\mathbb{Z}_2$ symmetries combine to form a dihedral group. How can we understand this from the 6d SCFT?

Here is where the 3-group structure of the 6d SCFT plays a role. Specifically, we have seen that the 6d SCFT contains a 3-group structure of the form: $\delta C = \mathcal{P} (B)$. Reducing it on the torus we get: $\delta A_C = 2A_{B_1}\cup A_{B_2}$, where we use $A_C$ for the background gauge field of the 4d $\mathbb{Z}_4$ symmetry coming from the 6d 2-form symmetry wrapping the torus and $A_{B_1}$, $A_{B_2}$ for the background gauge fields of the 4d $\mathbb{Z}_2$ symmetries coming from the 6d 1-form symmetry wrapping the two cycles of the torus, respectively. As only the $\mathbb{Z}_2$ subgroup of the $\mathbb{Z}_4$ symmetry survives in 4d, we need to set: $A_C = 2\hat{A}$, for $\hat{A}$ the background gauge field for the $\mathbb{Z}_2$ subgroup. Overall, we see that the 6d 3-group structure leads to the 4d extension: $\delta \hat{A} = A_{B_1}\cup A_{B_2}$, which as we have seen implies that the three $\mathbb{Z}_2$ symmetries combine to form the group $D_4$ such that the central element is the $\mathbb{Z}_2$ subgroup of the 2-form symmetry wrapping the torus. This precisely matches our observations in 4d.    

Finally, we consider the 1-form symmetries in 5d and their reduction. Due to the twist, we again only get the 1-form symmetries that are symmetric under the twist. These include the 1-form symmetry associated with the $SU(2)$ gauge group and the diagonal one associated with the four $E_1$ SCFTs. These form the $\mathbb{Z}_4$ 1-form symmetry of the 5d theory. Their reduction follows the discussion we had previously about the 4d reduction of the $E_1$ SCFT so we won't repeat it here. Like in the previous case discussed, the 4d reduction of these two symmetries should form an extension giving a $\mathbb{Z}_4$ 1-form symmetry also in 4d. The above account to two out of the three 1-form symmetries in 4d, leaving the last one associated with only a single $U(1)$ gauge group. Overall then, we find a $\mathbb{Z}_4 \times \mathbb{Z}_2$ 1-form symmetry group in 4d. This is exactly what we expect from 6d, where the $\mathbb{Z}_4$ part coming from the self-dual 2-form symmetry and the $\mathbb{Z}_2$ part coming from the 6d 1-form symmetry. 

 One subtle issue here is that the $\mathbb{Z}_2$ 1-form symmetry acting on an individual $U(1)$ group does not commute with the symmetry exchanging the 2 $U(1)$ gauge groups. This is weird as the former should be identified with the reduction of the 6d 1-form symmetry, while the latter is identified with the same symmetry wrapping the circle. A hint to the possible explanation of this is the fact that the two commute up to the diagonal 1-form symmetry, which is the subgroup of the $\mathbb{Z}_4$ 1-form symmetry. Now the 6d 3-group structure should also give the following 2-group relation in 4d: $\delta X_C = 2A_{B_1}\cup B$, coming from wrapping it on only one cycle of the torus. We interpret the observed non-commutativity as a result of this higher group structure. Indeed, if we set $X_C = 2\hat{B}$, that is we set the background field for the $\mathbb{Z}_2$ quotient to zero, the relation becomes $\delta \hat{B} = A_{B_1}\cup B$, and as argued in appendix \ref{App:D4}, this relation suggests that the symmetries associated with $A_{B_1}$ and $B$ commute only up to the symmetry associated with $\hat{B}$. This is precisely what we observe in 4d. This 2-group relation have also appeared in the previous 4d and 5d theories, and we expect it to have a similar interpretation, though in these cases there were also accidental 1-form symmetries complicating a precise matching. 

Overall, we see that we can match the symmetry properties of the 4d theory with those expected from the 6d SCFT reduces on a torus with a non-trivial Stiefel-Whitney class. Like the previous cases, the matching reveals the interesting interplay between symmetries of differing form degree and the relationship between them. Specifically, we have again seen how a higher group structure, here a 3-group structure, can reduce to a standard extension between 0-form symmetries. We have also seen how reducing a 2-group in the presence of a twist by one of its 0-form symmetries leads to the breaking of some of the 0-form symmetries involved in the 2-group. This has been observed already for continuous symmetries, and this example provides a realization of this for discrete symmetries. 

\section{Conclusions}
\label{sec:concl}

In this paper we have explored the behavior of generalized symmetries under dimensional reduction. We were particularly interested in cases where there is a non-trivial mixing between 0-form, 1-form and 2-form symmetries. We considered a set of simple examples, involving strongly coupled SCFTs, were the reduction can be followed and where most of the symmetries retain a non-trivial action in lower dimensions. This allows us to observe and illustrate various phenomena involving the interplay between dimensional reduction and generalized symmetries. 

More specifically, we have seen that a split extension between the 0 and 1-form symmetries reduces to a semi-direct product between the resulting 0-form symmetries. This leads to an enlarged symmetry group in lower dimensions, that keeps growing as the dimension is reduced. In the example of the 6d $so(8)$ SCFT, it provides an additional perspective on why the lower dimensional theory has the shape of a $D_4$ Dynkin diagram. Specifically, the 4d theory has a $W(D_4)$ 0-form global symmetry, which is not present in the original 6d SCFT. As such, we might naively deem it as accidental. However, we see that in fact it entirely originates in symmetries of the 6d SCFTs, but many of them are higher form symmetries that reduce to 0-form symmetries. The higher group structure mixing these symmetries then leads to the extensions building the group $W(D_4)$.    

Similarly, it is possible to reduce a 2-group to another type of extension, an anomaly. This is most readily apparent from the symmetry TQFT, where the 2-group term is similar to an anomaly involving a gauged $d-3$ form symmetry (associated with the symmetry dual to the involved 1-form symmetry) and the extension class. As such when the extension class integrated over the circle results in the background field for a discrete symmetry, said symmetry should be broken by the 2-group. We have seen examples of this in the case of the outer automorphism symmetry of the 6d $so(2n)$ SCFT when it is reduced with a non-trivial SW class, and there are seemingly related phenomena in 2-group structure for continuous symmetries \cite{Nardoni:2024sos}. It would be interesting to better understand this.

Matching global symmetries also allows us to put new tests on the dualities between the higher dimensional theory reduced on a compact space, and the proposed purely lower dimensional description. Even-though such relations are often marred with accidental or trivially acting symmetries, we have proposed a reasonable mapping of the symmetries between the different descriptions, which often also made use of various generalized symmetry properties of the involved theories. The overall matching then provides an additional consistency check of the many generalized symmetry properties and interrelations of the involved theories. Despite that, there were also various properties whose precise matching requires further study, as well as incidents of both accidental symmetries and trivially acting ones. It is clear that further work on this front is required.

There are various additional angles one can consider. First, we can consider more general reductions of the 6d $so$ SCFTs considered here, like ones involving also discrete symmetry twists in 0 or 1-form symmetries. Second, we can also consider the dimensional reduction of other 6d SCFTs, or even other strongly coupled theories like 5d SCFTs, that have interesting discrete symmetry structures. Alternatively, we can consider interesting 4d theories or 3d theories, with a known higher dimensional construction, that exhibit intricate discrete symmetry structure. Based on the above example, it is possible for such a structure to originate from generalized symmetry structure of the 6d SCFT, so this may provide a novel approach to exploring generalized symmetries of 6d or 5d SCFTs.  

Finally, it will also be interesting to understand the relation between not just the symmetries themselves, but also the anomalies involving them. We have already seen that the 5d theory has several 't Hooft anomalies involving discrete symmetries, and it is interesting if these can be matched to anomalies of the 6d SCFT. This would them provide an additional check and the proposed relation between the theories and symmetries. It may also be helpful to thoroughly study the anomalies of the involved theories. This is as these examples have a relative lack of trivially acting symmetries in lower dimensions, and it is tempting to associate this with the potential presence of 't Hooft anomalies. It will be intriguing to explore this further.    

\section*{Acknowledgments}
We would like to thank Emily Nardoni, Shlomo Razamat, Matteo Sacchi, Orr Sela and Yunqin Zheng for helpful discussions. GZ is partially supported by the Israel Science Foundation under grant no. 759/23.

\begin{appendix}

\section{Illustrating examples in 5d}
\label{App:5dSU4}

\begin{figure}
\center
\includegraphics[width=0.75\textwidth]{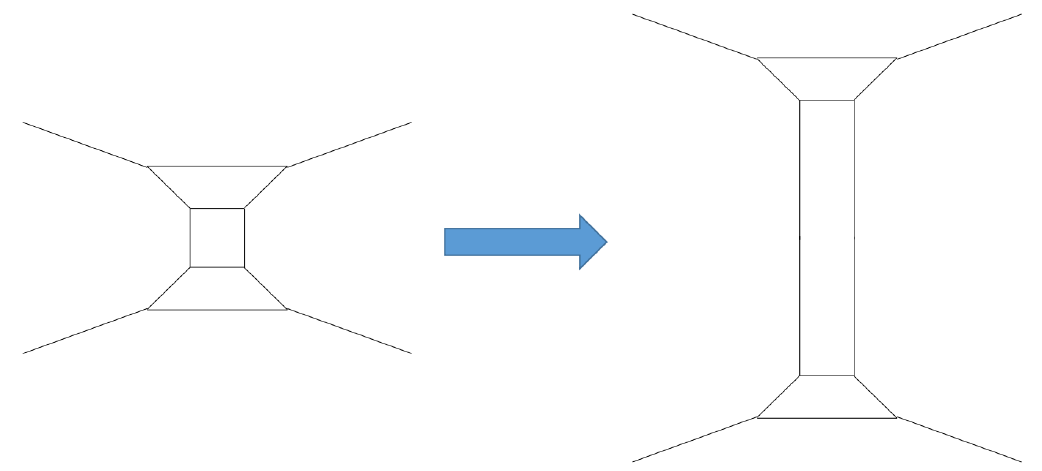} 
\caption{On the left is the web describing the 5d SCFT UV completing an $SU(4)_0$ gauge theory. One can see that it consists of 4 parallel D5-branes on which lives an $SU(4)$ gauge theory, while the invariance of the web under reflections implies that the Chern-Simons term is zero. Stretching the D5-branes corresponds to making the $SU(4)$ gauge coupling small and leads to the effective description as an $SU(4)_0$ gauge theory. However, in the opposite limit, depicted on the right, the theory has an effective description as a weakly coupled $SU(2)$ gauge group gauging the $SO(3)$ flavor symmetry of two $E_1$ SCFTs.}
\label{WebSU4}
\end{figure}

Above we discussed the anomalies and symmetries of the gauging of four $E_1$ 5d SCFTs with an $SU(2)$ gauge group and compared them with the expectations from the reduction of the 6d $so(8)$ SCFT. It is instructive to consider a similar but simpler example where we gauge just two $E_1$ SCFTs. The resulting theories then describe a 5d SCFT which is the UV completion of a pure $SU(4)_0$ gauge theory (as before the subscript denotes the Chern-Simons term), see figure \ref{WebSU4}. It is then interesting to compare the expectations between the two low-energy descriptions of the 5d SCFT. 

We begin with listing the properties of the Lagrangian theory. Its continuous flavor symmetries are merely a $U(1)$ that appear as the topological symmetry in the gauge theory. This remains a $U(1)$ also in the SCFT. Additionally, the theory has a $\mathbb{Z}_4$ 1-form symmetry and a $\mathbb{Z}_2$ discrete symmetry, associated with charge conjugation.     

At the gauge theory level, there is a mixed anomaly involving the 1-form symmetry and the $U(1)$ flavor symmetry. It is given by \cite{BenettiGenolini:2020doj}:

\be
\frac{3\pi i}{4} \int \tilde{C}_1 \cup P (B) ,
\ee
with $B$ the background gauge field for the $\mathbb{Z}_4$ 1-form symmetry and $\tilde{C}_1$ the mod $8$ reduction of the first Chern class of the $U(1)$. Note that this mixed anomaly only affects the entire $\mathbb{Z}_4$ and not its $\mathbb{Z}_2$ subgroup. Specifically, if we set $B = \hat{B} + 2 B_1$ (with $\hat{B}$ and $B_1$ $\mathbb{Z}_2$ valued forms), then we have that:

\bea \label{Anomz4toz2}
&& \frac{3\pi i}{4} \int \tilde{C}_1 \cup P (B) \rightarrow \frac{3\pi i}{4} \int \tilde{C}_1 \cup ( P (\hat{B}) + 4 P (B_1) + 4 \hat{B}\cup B_1 ) \\ \nonumber && = \frac{3\pi i}{4} \int \tilde{C}_1 \cup P (\hat{B}) + \pi i \int \tilde{C}_1 \cup \hat{B}\cup B_1 ,
\eea
where we have used the fact that $P (B_1)$ is even on spin manifolds. Note that the above anomaly always contain $\hat{B}$. Finally, we note that there could be other anomalies, also involving charge conjugation and the R-symmetry, but we shall not study this here.

Next we turn to the dual description as an $SU(2)$ gauging of two $E_1$ SCFTs. Here there is also a $U(1)$ flavor symmetry, associated with the topological symmetry of the $SU(2)$ gauge group. This should map to the $U(1)$ of the SCFT. The theory also has a $\mathbb{Z}_2$ discrete symmetry, associated with quiver reflection, which is expected to map to charge conjugation. Finally, the theory has 3 $\mathbb{Z}_2$ 1-form symmetries, 2 coming from the two $E_1$ SCFTs, and one coming from the $SU(2)$ gauge group\footnote{Here it is important that the theta angle of the $SU(2)$ is $\theta=0$, so the 1-form symmetry is present also in the 5d SCFT. The value of the $\theta$ angle in this case an be inferred from a similar duality considered in \cite{Bergman:2013aca}.}. These should somehow reproduce the $\mathbb{Z}_4$ 1-form symmetry of the 5d SCFT.

This works in a very similar way to the 6d $so(8)$ case we analyzed, though it is much simpler here as we only have two $E_1$ SCFTs. The idea again is that the presence of the 2-group structure of the $E_1$ SCFTs leads to the extension. We can again explore this using a mock symmetry TQFT:

\be
\pi i \int \hat{c} \cup \delta \hat{B} + c_1 \cup \delta B_1 + c_2 \cup \delta B_2 + c_1 \cup \beta \hat{B} + c_2 \cup \beta \hat{B} ,
\ee    
where $\hat{B}$ is the background field for the $SU(2)$ 1-form symmetry, $B_{1,2}$ the background fields for the 1-form symmetries of the two $E_1$ SCFTs, and $\hat{c}, c_{1,2}$ are the conjugate 3-form gauge fields. We can now make the following change of variables:

\bea
\begin{pmatrix}
  c^+_1 \\
  c_2
\end{pmatrix} = \begin{pmatrix}
  1 & 1 \\
  0 & 1
\end{pmatrix}\begin{pmatrix}
  c_1 \\
  c_2
\end{pmatrix} \; , \; \begin{pmatrix}
  B_1 \\
  B^+_2
\end{pmatrix} \rightarrow \begin{pmatrix}
  1 & 0 \\
  1 & 1
\end{pmatrix}\begin{pmatrix}
  B_1 \\
  B_2
\end{pmatrix} .
\eea

Note that this is a well defined change of variables between fields with integer coefficients. Using this we can rewrite the symmetry TQFT has:

\be
\pi i \int \hat{c} \cup \delta \hat{B} + c^+_1 \cup (\delta B_1 + \beta \hat{B}) + c_2 \cup \delta B^+_2 .
\ee 

We note that we can combine $B_1$ and $\hat{B}$ into the the $\mathbb{Z}_4$ gauge field $X = \hat{B} + 2 B_1$. The closeness of $X$ then forces the condition $\delta B_1 + \beta \hat{B} = 0$, coming from the $E_1$ 2-group structure. We therefore conclude that $X$ should match the $\mathbb{Z}_4$ 1-form symmetry of the 5d SCFT, while $B^+_2$ is an accidental $\mathbb{Z}_2$ 1-form symmetry. 

Next we turn to comparing the anomalies. There are two sources for anomalies in the $SU(2)$ theory. One involves the fractional instantons of the $SU(2)$ that arise when its 1-form symmetry is gauged. This is similar to the anomaly on the $SU(4)_0$ side, and takes the form:

\be
\frac{\pi i}{2}\int \tilde{C}'_1 \cup P(\hat{B}) ,
\ee  
where $\tilde{C}'_1$ is the first Chern class of the $U(1)$ instanton symmetry associated with the $SU(2)$. Comparing this with \eqref{Anomz4toz2}, we need to identify $2 \tilde{C}'_1 = -\tilde{C}_1$ to match the anomaly to the first term of \eqref{Anomz4toz2}. The minus sign just represents charge conjugation, but the factor of 2 is more interesting. It suggests that the 1-instanton on the $SU(4)$ side is mapped to the 2-instanton on the $SU(2)$ side. Note that with this identification, the second term should trivialize.

We see then that we can identify this anomaly between the two theories. It might be possible to observe the additional terms expected from \eqref{Anomz4toz2}, but we shall not pursue it here.

This leads us to the second source of anomaly, the one from the $E_1$ SCFTs themselves. These have the form of:

\be
\frac{\pi i}{2}\int \hat{B} \cup ( P(B_1) + P(B_2) ) .
\ee 

To better understand these anomalies, we first want to convert to $B_1$ and $B^+_2$, which we can do by setting $B_2 = B^+_2 + B_1$. Expanding then gives:

\be
\frac{\pi i}{2}\int \hat{B} \cup ( P(B_1) + P(B^+_2 + B_1) ) = \frac{\pi i}{2}\int \hat{B} \cup ( P(B^+_2) + 2 B^+_2 \cup B_1).
\ee 

The first thing we note is that all the above anomalies contain $B^+_2$, and so involve the accidental symmetries. As such, these probably do not arise in the SCFT.    

\subsection{Generalization}

It is interesting to consider the generalization of the above relation for the $SU(4)_0$ SCFT. Specifically, consider the 5d gauge theory $SU(n+m)_{n-m}$. This gauge theory can be UV completed by a 5d SCFT. In the specific case of $SU(n)_{\pm n}$, said SCFT has an enhanced $su(2)$ flavor symmetry (whose Cartan is the topological symmetry of the gauge theory) \cite{Bergman:2013aca}. It also has a $\mathbb{Z}_n$ 1-form symmetry. At the group level, the global symmetry is thought to be $SO(3)$ and when $n$ is even, it combines with the 1-form symmetry to form a 2-group: $\delta B = \beta_n w_2$ (where here we use $\beta_n$ for the Bockstein associated with the extension $0 \rightarrow \mathbb{Z}_n \rightarrow \mathbb{Z}_{2n} \rightarrow \mathbb{Z}_2 \rightarrow 0$) \cite{Apruzzi:2021vcu}. For generic values of $n$ and $m$, the symmetry remains a $U(1)$. We also still have a $\mathbb{Z}_{gcd(n+m,n-m)}$ 1-form symmetry.

\begin{figure}
\center
\includegraphics[width=0.75\textwidth]{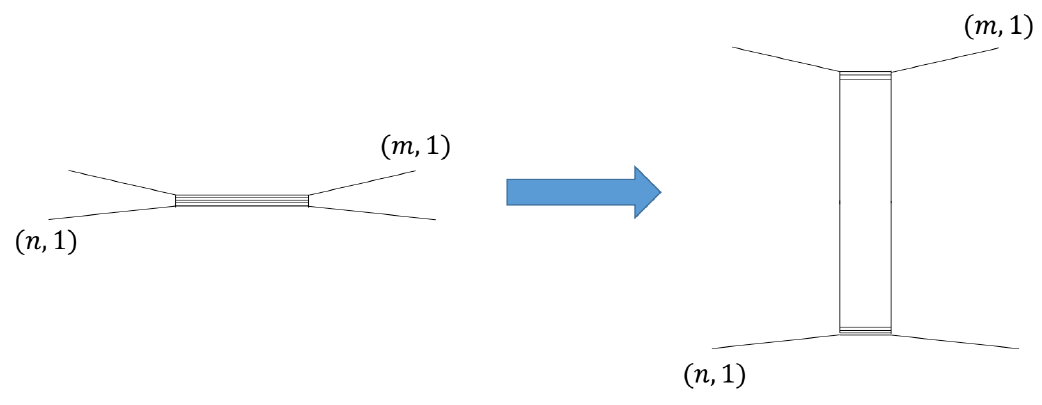} 
\caption{The web for the 5d SCFT UV completing an $SU(n+m)_{n-m}$ gauge theory. Here the numbers in parenthesis denote the 5-brane charges of some of the external legs. On the left is the web in the frame manifesting its 5d gauge theory description, where we separated the web such that we have $n+m$ D5-branes stretching between the two edges. In the opposite limit, depicted on the right, the theory can be effectively described by an $SU(2)$ gauge theory gauging the $SO(3)$ global symmetry of two 5d SCFTs, one UV completing an $SU(n)_{n}$ gauge theory, while the other UV completing an $SU(m)_{-m}$ one.}
\label{WebSUnm}
\end{figure}

Now consider the $SU(n+m)_{n-m}$ SCFT, whose deformed brane web is depicted in figure \ref{WebSUnm}. When given a positive mass deformation in its $U(1)$ flavor symmetry, it flows to the $SU(n+m)_{n-m}$ gauge theory. However, what happens if we give it a mass deformation with a negative sign? In the $SU(n)_{\pm n}$ case, we still get the same gauge theory, due to the Weyl group of the enhanced $SO(3)$, but in the generic case we get a different IR theory. Using brane webs, it is possible to show that the IR theory in this case is given by an $SU(2)$ gauging of two 5d SCFTs, one UV completing an $SU(n)_{n}$ gauge theory, and another UV completing an $SU(m)_{-m}$ one (see figure \ref{WebSUnm}). It is then interesting how the global symmetry of the SCFT is manifested in the low-energy theory.

The case of most interest is when $n=m$, and we have an $SU(2n)_0$ gauge theory, and its associated UV completion. The global symmetries in this case consist of a continuous $U(1)$ symmetry, a $\mathbb{Z}_{2n}$ 1-form symmetry and a $\mathbb{Z}_2$ 0-form symmetry manifested as charge conjugation on the gauge theory side. Here the $U(1)$ just maps to the instanton $U(1)$ of the $SU(2)$ gauge group. The charge conjugation symmetry should map to the quiver reflection symmetry. This leaves the $\mathbb{Z}_{2n}$ 1-form symmetry. Note that on the dual side, we observe only a $\mathbb{Z}_{2} \times \mathbb{Z}_{n} \times \mathbb{Z}_{n}$ 1-form symmetry. For $n$ odd there is no problem as $\mathbb{Z}_{2n} = \mathbb{Z}_{2} \times \mathbb{Z}_{n}$, so the symmetries can be matched, though we are again forced to interpret a $\mathbb{Z}_{n}$ 1-form symmetry as an accidental enhancement.

For $n$ even however, there is an apparent mismatch as $\mathbb{Z}_{2n}$ and $\mathbb{Z}_{2} \times \mathbb{Z}_{n}$ are different groups for even $n$. The resolution of this mismatch is the additional 2-group structure. As before, the 2-group structure of the $SU(n)_{\pm n}$ SCFT, $\delta B_1 = \beta_n w_2$, leads to the extension $\delta B_1 = \beta_n \hat{B}$ once the $SO(3)$ global symmetry is gauged. This then leads to the non-trivial extension turning $\mathbb{Z}_{2} \times \mathbb{Z}_{n}$ into $\mathbb{Z}_{2n}$. Overall, we see that the 2-group structure is necessary to match the 1-form symmetry precisely for $n$ even, as expected from \cite{Apruzzi:2021vcu}.

It is instructive to see this also from the background fields. As before, we can study this using the symmetry TFT:

\be
\pi i \int \hat{c} \cup \delta \hat{B} + \frac{\pi i}{n} \int c_1 \cup \delta B_1 + c_2 \cup \delta B_2 + c_1 \cup \beta_n \hat{B} + c_2 \cup \beta_n \hat{B} .
\ee  

Specifically, we can again rewrite this as:

\be
\pi i \int \hat{c} \cup \delta \hat{B} + \frac{\pi i}{n} \int c^+_1 \cup (\delta B_1 + \beta_n \hat{B}) + c_2 \cup \delta B^-_2 ,
\ee
where $c^+_1 = c_1 + c_2$, and $B^-_2 = B_2 - B_1$. This indeed leads to the extension $\delta B_1 + \beta_n \hat{B} = 0$. Note that this suggests that the charges under the $\mathbb{Z}_n$ that participate in the extension are given by $q_2 - q_1$ \footnote{This is a bit subtle. To illustrate it, consider a $U(1)_1 \times U(1)_2$ symmetry. Say we now switch to the combination $U(1)_a = U(1)_1 + U(1)_2$ and $U(1)_b = U(1)_2$. We can ask how are the charges under $U(1)_a$ and $U(1)_b$ related to those under $U(1)_1$ and $U(1)_2$. The answer in this case is: $q_a = q_1$, $q_b = q_2 - q_1$. This is as $U(1)_1$ only appears in $U(1)_a$ and as such the $U(1)_1$ charge automatically gives the $U(1)_a$ charge. However, $U(1)_2$ appears in both, but appears symmetrically with $U(1)_1$ in $U(1)_a$ and so their difference is the $U(1)_2$ charge. More generally under the symmetry transformation: $V'_s = M^T V_s$, for $V_s$ a vector of symmetries and $M$ a matrix, the vector of charges transforms as: $V'_q = M^{-1}V_q$.}. This is interesting as under the charge conjugation of $\mathbb{Z}_{2n}$, the $\mathbb{Z}_n$ subgroup is charge conjugated. Note that under quiver reflection we have that: $q_2 - q_1 \rightarrow - (q_2 - q_1)$, which is exactly the action of charge conjugation. Therefore, we see that the action of quiver reflection is indeed consistent with it mapped to charge conjugation.

It is also possible to consider more general values of $n$ and $m$. When $gcd(n+m,n-m)=1$ the 1-form symmetry of the 5d SCFT is trivial and the 1-form symmetries of the 5d theory can be attributed to accidental enhancements. However, when $gcd(n+m,n-m)\neq 1$ the 5d SCFT has a non-trivial 1-form symmetry, which should also be present in the low-energy 5d theory. Naturally, if $gcd(m,n) = l$ we would have at least a $\mathbb{Z}_{l}$ 1-form symmetry. It is easy to see that this symmetry would also be present in the 5d theory as we have a $\mathbb{Z}_m$ and $\mathbb{Z}_n$ 1-form symmetries. 

The interesting case occurs when $gcd(n+m,n-m)>gcd(m,n)$, in which case the 1-form symmetry of the 5d SCFT may not be contained in $\mathbb{Z}_m \times \mathbb{Z}_n$. We can write $n+m = x l$, $n-m = y l$, with $gcd(x,y)=1$. This gives $2n=(x+y)l$, $2m=(x-y)l$, from which we see that $gcd(n+m,n-m)>gcd(m,n)$ can only happen if $l$ is even and only one of $x$, $y$ is odd. In this case the 1-form symmetry of the 5d SCFT is $\mathbb{Z}_{2r}$, while that of the 5d theory is $\mathbb{Z}_{(x-y)r} \times \mathbb{Z}_2 \times \mathbb{Z}_{(x+y)r}$, where we have taken $l=2r$.

If $r$ is odd, then we can write $\mathbb{Z}_{2r} = \mathbb{Z}_2 \times \mathbb{Z}_r$, and the symmetries again immediately match. Say now that $r=2^p q$ for $q$ odd, then we can write $\mathbb{Z}_{2r} = \mathbb{Z}_{2^{p+1}} \times \mathbb{Z}_q$ ,$\mathbb{Z}_{(x-y)r} = \mathbb{Z}_{(x-y)q} \times \mathbb{Z}_{2^p}$, $\mathbb{Z}_{(x+y)r} = \mathbb{Z}_{(x+y)q} \times \mathbb{Z}_{2^p}$, where we used the fact that $x\pm y$ is odd. Due to the 2-group of the $E_1$ SCFT, we get the extension of $\mathbb{Z}_2$ by $\mathbb{Z}_{2^p}$ to $\mathbb{Z}_{2^{p+1}}$, so we again see that the 1-form symmetry of the 5d SCFT are present in the 5d theory. Overall, we see that the mismatch is at best by an extension of a $\mathbb{Z}_2$, which is then taken care of by the 2-group structure of the $E_1$ SCFT. 

\section{The group $D_4$}
\label{App:D4}

Here we collect some properties of the group $D_4$ that might be useful. The group has $8$ elements and is generated by two generators $\alpha$, $x$ subject to the relations: $x^2 = \alpha^4 = 1$, $x \alpha x = \alpha^{-1}$. Note that the element $\alpha^2$ is special as it commutes with all elements of the group. Additionally, the elements ${1,\alpha,\alpha^2,\alpha^3}$ form a $\mathbb{Z}_4$ normal subgroup, and the elements ${1,x,\alpha^2,\alpha^2 x}$ and ${1,x \alpha,\alpha x,\alpha^2}$ form a $\mathbb{Z}_2 \times \mathbb{Z}_2$ normal subgroup.

The group has only one faithful irreducible representation which is 2-dimensional. Additionally, it has 4 1d irreducible representation, which are essentially representations of its $\mathbb{Z}_2 \times \mathbb{Z}_2$ quotient (under ${1,\alpha^2}$), given by $(x,\alpha)\rightarrow (\pm 1, \pm 1)$. The group can be described by the semi-direct products, $\mathbb{Z}_4 \rtimes \mathbb{Z}_2$ and $(\mathbb{Z}_2 \times \mathbb{Z}_2)\rtimes \mathbb{Z}_2$.  

\subsection*{Central extension and background fields}

In addition to its presentation as a semi-direct product, the group $D_4$ can also be represented as a central extension of $\mathbb{Z}_2 \times \mathbb{Z}_2$ by $\mathbb{Z}_2$. This representation plays some role in the discussion above so we shall consider it here. Specifically, central extensions are classified by the group cohomology element $H^2(Q,H)$. For $Q=\mathbb{Z}_2 \times \mathbb{Z}_2$, $H = \mathbb{Z}_2$ we have that $H^2(\mathbb{Z}_2 \times \mathbb{Z}_2, \mathbb{Z}_2) = \mathbb{Z}_2 \times \mathbb{Z}_2 \times \mathbb{Z}_2$, so that there are eight different possible extensions. These include the direct product, $\mathbb{Z}_2 \times \mathbb{Z}_4$ (which appears three times), the group $D_4$ (which also appear three times), and the quaternion group $Q_4$. 

It is interesting to see how the extension manifests at the level of the background fields. Specifically, we can associate a $\mathbb{Z}_2$ valued 1-form for each symmetry, where we shall use $\hat{A}$ for the $\mathbb{Z}_2$ and $A_1$, $A_2$ for the $\mathbb{Z}_2 \times \mathbb{Z}_2$. The extension of $\mathbb{Z}_2$ by $\mathbb{Z}_2$ to $\mathbb{Z}_4$ is described by the background fields through the relation: $\delta \hat{A} = \beta A_1 = A_1 \cup A_1$. This should then also describe the extension of $\mathbb{Z}_2 \times \mathbb{Z}_2$ to $\mathbb{Z}_2 \times \mathbb{Z}_4$. Note that here we could have also chosen to extend $A_2$ or $A_1+A_2$ to $\mathbb{Z}_4$ instead. This would then be described by the relation: $\delta \hat{A} = \beta A_2 = A_2 \cup A_2$ and $\delta \hat{A} = \beta (A_1 + A_2) = (A_1 + A_2) \cup (A_1 + A_2)$, respectively.

More generally, we can consider the most general 2-from we can build from $A_1$ and $A_2$, and look at the relation:

\be
\delta \hat{A} = \kappa A_1 \cup A_1 + \lambda A_1 \cup A_2 + \mu A_2 \cup A_2 ,
\ee
where $\alpha$, $\beta$ and $\gamma$ are $\mathbb{Z}_2$ valued coefficients. Overall, there are $8$ such terms, forming the group $\mathbb{Z}_2 \times \mathbb{Z}_2 \times \mathbb{Z}_2$. These in turn are in one-to-one correspondence with extensions of $\mathbb{Z}_2 \times \mathbb{Z}_2$ by $\mathbb{Z}_2$. Indeed, we have seen that the case of $(\kappa,\lambda,\mu) = (0,0,0)$ corresponds to the direct product, while the cases $(1,0,0)$, $(0,0,1)$ and $(1,0,1)$ correspond to the $\mathbb{Z}_2 \times \mathbb{Z}_4$ case. The cases $(0,1,0)$, $(1,1,0)$ and $(0,1,1)$ then correspond to $D_4$ (one can see that the three rotate to one another under the transformations $A_1 \rightarrow A_1 + A_2$, and $A_2 \rightarrow A_1 + A_2$), while $(1,1,1)$ gives $Q_4$, see \cite{Bergman:2024its}.

At the level of the generators, the group is then given by:

\be
a^2_1 = \hat{a}^{\kappa} , a^2_2 = \hat{a}^{\mu} , a_1 a_2 a^{-1}_1 a^{-1}_2 = \hat{a}^{\lambda} .
\ee

This can be understood as follows. The terms with $\kappa$ and $\mu$ describe the standard $\mathbb{Z}_4$ extension we have seen that originates from $\delta \hat{A} =\beta A_i = A_i \cup A_i$. The interesting term is the one coming from $\lambda$. Note that this leads to the non-commutativity of the group. In terms of the background fields we would write (schematically):

\be \label{CRD4}
e^{\int A_1} e^{\int A_2} e^{-\int A_1} e^{-\int A_2} = e^{\lambda\int \hat{A}} .
\ee

When $\lambda=1$, this cannot be satisfied if the $A_i$ commute, and as such, we should take them to be matrices. The easiest solution is to take $[A_1,A_2] = \lambda \hat{A}$, $[\hat{A},A_1] = [\hat{A},A_2] = 0$ in which case we can use the Baker-Campbell-Housdorff formula to rewrite \eqref{CRD4} as: 

\be
e^{\int A_1} e^{\int A_2} e^{-\int A_1} e^{-\int A_2} \rightarrow e^{\int A_1 + A_2 + \frac{1}{2}\lambda \hat{A}}e^{\int -A_1 - A_2 + \frac{1}{2}\lambda \hat{A}} \rightarrow e^{\int\lambda \hat{A}} ,
\ee
as required.

As we do in non-abelian gauge theories, we can then formally write the connection as: $\bold{A} = A_1 T_1 + A_2 T_2 + \hat{A} \hat{T}$, where now $A_i$, $\hat{A}$ are commuting $\mathbb{Z}_2$ valued 1-forms and $T_1$, $T_2$ and $\hat{T}$ are matrices obeying $[T_1,T_2] = \lambda \hat{T}$, $[T_1,\hat{T}] = [T_2,\hat{T}] = 0$. This then ensures that the commutation relation \eqref{CRD4} is obeyed.

We can then think of $\bold{A}$ as the gauge field for the discrete symmetry. However, since the symmetry is discrete the gauge field must be flat, that is $D \bold{A} = \delta \bold{A} + \frac{1}{2}[\bold{A},\bold{A}] = 0$, where here we have used the covariant derivative as $\bold{A}$ is non-commuting. When written in components, we get the condition: $\delta A_1 T_1 + \delta A_2 T_2 + \delta \hat{A} \hat{T} + \lambda A_1 \cup A_2 \hat{T} = 0$. Comparing coefficients, we then get the conditions: $\delta A_1 = 0$ , $\delta A_2 = 0$ , $\delta \hat{A} = \lambda A_1 \cup A_2$, which are indeed the required relations when $\kappa=\mu=0$.

There is another way to argue for the non-commutativity using the gauge transformation of the background fields. For this we introduce the following 3d symmetry TQFT studied in \cite{Kaidi:2023maf}:

\be
\pi \int \hat{a} \cup\delta a + \hat{b} \cup\delta b + \hat{c} \cup\delta c + a \cup b \cup c .
\ee

This symmetry TQFT describes an extension of the form: $\delta \hat{a} = b \cup c$, assuming $\hat{a}$, $b$ and $c$ are given Dirichlet boundary conditions. Here all fields are 1-forms, but it is straightforward to generalize to other form degrees, with minor modifications to the proceeding analysis. For instance, we can turn it into a symmetry TQFT describing a similar extension for a 4d theory by taking $a$, $\hat{b}$ and $\hat{c}$ to be 3-form gauge fields. We also take all fields to be $\mathbb{Z}_2$ cochains. 

The fields obey the following gauge symmetries:

\bea
& & a \rightarrow a + \delta \alpha \; , \; b \rightarrow b + \delta \beta \; , \; c \rightarrow c + \delta \gamma \; , \nonumber \\
& & \hat{a} \rightarrow \hat{a} + \delta \hat{\alpha} + \beta c - \gamma b \; , \nonumber \\
& & \hat{b} \rightarrow \hat{b} + \delta \hat{\beta} + \gamma a - \alpha c \; , \nonumber \\
& & \hat{c} \rightarrow \hat{c} + \delta \hat{\gamma} + \alpha b - \beta a \; . \label{STFTgt}
\eea

Now consider performing the following sequence of gauge transformations: $a \rightarrow a + \delta \alpha$,  $b \rightarrow b + \delta \beta$, $a \rightarrow a - \delta \alpha$,  $b \rightarrow b - \delta \beta$. Note that the final transformation just takes $a \rightarrow a$ and $b \rightarrow b$, but due to the mixed terms in \eqref{STFTgt}, we have that $\hat{a} \rightarrow \hat{a} + \beta \cup \delta \alpha - \alpha \cup\delta \beta = \hat{a} + \delta (\alpha \cup \beta)$. We see then that the relation $\delta \hat{a} = b \cup c$ implies that the gauge transformations of $b$ and $c$ don't commute, and yield a gauge transformation of $\hat{a}$.

The above analysis can be immediately generalized to other cases. For instance, consider the case where we have the relation: $\delta \hat{B} = A \cup B$, where $A$ is a 1-form but $B$ and $\hat{B}$ are 2-forms. A similar analysis yields that the gauge transformations of $A$ and $B$ do not commute, rather their commutator resulting in the gauge transformation of $\hat{B}$. 

\section{Weyl group of $D_4$}
\label{App:WD4}

Consider the Weyl group of $so(8)$. A convenient way to describe it is as the residual gauge symmetry after diagonalizing an $so(8)$ adjoint matrix. The matrix then will take the form: $M=diag(\pm x_1, \pm x_2, \pm x_3, \pm x_4)$, and the Weyl group is generated by the permutations of the four $x_i$'s, as well as the transformations taking $x_i \rightarrow -x_i$ for an even number of the $x_i$'s\footnote{The transformation taking $x_i \rightarrow -x_i$ for just one of the $x_i$'s is an outer automorphism rather than an inner one.}. A convenient representation of it can now be given by $4\times 4$ matrices implementing the permutations and double reflections.   

Note that the element $-I$, corresponding to the transformation $x_i \rightarrow -x_i$ for all the $x_i$'s, is central and so we can quotient by it. We shall denote the resulting group by $W(D_4)/\mathbb{Z}_2$. Consider now the following collection of matrices:

\bea \label{WD4Z2q}
& & \begin{pmatrix}
  -1 & 0 & 0 & 0 \\
  0 & -1 & 0 & 0 \\
  0 & 0 & 1 & 0 \\
  0 & 0 & 0 & 1
\end{pmatrix} \; , \; \begin{pmatrix}
   -1 & 0 & 0 & 0 \\
  0 & 1 & 0 & 0 \\
  0 & 0 & -1 & 0 \\
  0 & 0 & 0 & 1
\end{pmatrix} \; , \; \begin{pmatrix}
   -1 & 0 & 0 & 0 \\
  0 & 1 & 0 & 0 \\
  0 & 0 & 1 & 0 \\
  0 & 0 & 0 & -1
\end{pmatrix} \; , \\ \nonumber
& & \begin{pmatrix}
  0 & 1 & 0 & 0 \\
  1 & 0 & 0 & 0 \\
  0 & 0 & 0 & 1 \\
  0 & 0 & 1 & 0
\end{pmatrix} \; , \; \begin{pmatrix}
  0 & 0 & 1 & 0 \\
  0 & 0 & 0 & 1 \\
  1 & 0 & 0 & 0 \\
  0 & 1 & 0 & 0
\end{pmatrix} \; , \; \begin{pmatrix}
  0 & 0 & 0 & 1 \\
  0 & 0 & 1 & 0 \\
  0 & 1 & 0 & 0 \\
  1 & 0 & 0 & 0
\end{pmatrix}.
\eea

These generate a $\mathbb{Z}^4_2$ subgroup of $W(D_4)/\mathbb{Z}_2$, where we use the fact that $M=-M$ since $-I$ was quotiented out. Here the first three matrices form the three non-trivial element of one $\mathbb{Z}^2_2$, and the next three matrices form the three non-trivial element of the other $\mathbb{Z}^2_2$. 

Now consider the following two additional elements:

\bea
& & \begin{pmatrix}
  1 & 0 & 0 & 0 \\
  0 & 0 & 1 & 0 \\
  0 & 1 & 0 & 0 \\
  0 & 0 & 0 & 1
\end{pmatrix} \; , \; \begin{pmatrix}
   1 & 0 & 0 & 0 \\
  0 & 0 & 1 & 0 \\
  0 & 0 & 0 & 1 \\
  0 & 1 & 0 & 0
\end{pmatrix} .
\eea

These generate an $S_3$ subgroup of $W(D_4)/\mathbb{Z}_2$, acting by the permutation of the coordinates $x_2-x_4$. Furthermore, note that these act on the matrices in \eqref{WD4Z2q} by permuting the three upper and lower matrices among themselves. Therefore, the group formed by all the matrices should be equivalent to $\mathbb{Z}^4_2 \rtimes S_3$, where the $S_3$ acts by the simultaneous permutation of the three $\mathbb{Z}_2$ subgroups of both pairs of $\mathbb{Z}^2_2$ groups. Moreover, this group must be equivalent to $W(D_4)/\mathbb{Z}_2$. This just follows as the order of both $W(D_4)/\mathbb{Z}_2$ and $\mathbb{Z}^4_2 \rtimes S_3$ is $96$. Thus, we conclude that $W(D_4)/\mathbb{Z}_2 = \mathbb{Z}^4_2 \rtimes S_3$.

Finally, it is instructive to consider the extension to the full $W(D_4)$. For this we first consider the extension by $-I$ of the $\mathbb{Z}^4_2$ subgroup of $W(D_4)/\mathbb{Z}_2$ given by the matrices \eqref{WD4Z2q}. To analyze it we pick four generators of $\mathbb{Z}^4_2$, which we take to be:

 \bea \label{Basis}
& & R_1 = \begin{pmatrix}
  -1 & 0 & 0 & 0 \\
  0 & -1 & 0 & 0 \\
  0 & 0 & 1 & 0 \\
  0 & 0 & 0 & 1
\end{pmatrix} \; , \; R_2 = \begin{pmatrix}
   -1 & 0 & 0 & 0 \\
  0 & 1 & 0 & 0 \\
  0 & 0 & 1 & 0 \\
  0 & 0 & 0 & -1
\end{pmatrix} \; , \\ \nonumber
& & E_2 = \begin{pmatrix}
  0 & 1 & 0 & 0 \\
  1 & 0 & 0 & 0 \\
  0 & 0 & 0 & 1 \\
  0 & 0 & 1 & 0
\end{pmatrix} \; , E_1 = \; \begin{pmatrix}
  0 & 0 & 0 & 1 \\
  0 & 0 & 1 & 0 \\
  0 & 1 & 0 & 0 \\
  1 & 0 & 0 & 0
\end{pmatrix}.
\eea

We note that the extension is now described by modifying the relations of $\mathbb{Z}^4_2$ as: $E_1 R_1 E_1 R_1 = -I$, $E_2 R_2 E_2 R_2 = -I$, with the rest of the relations unchanged. This suggests that the extension is a non-trivial central extension. The resulting group appears to be the extraspecial group $2^{1+4}_+$. It is apparent that the full $W(D_4)$ is then given by the semi-direct product of this group with the $S_3$, that is $W(D_4) = 2^{1+4}_+ \rtimes S_3$. Indeed, it is known that $W(D_4) = \mathbb{Z}^3_2 \rtimes S_4 = 2^{1+4}_+ \rtimes S_3$

The extension of $W(D_4)/\mathbb{Z}_2$ to $W(D_4)$ can then be described by background fields, similarly to the extension of $\mathbb{Z}_2 \times \mathbb{Z}_2$ to $D_4$ discussed in appendix \ref{App:D4}. Specifically, we introduce the $\mathbb{Z}_2$ background fields: $A_c$, $A_{R_1}$, $A_{R_2}$, $A_{E_1}$, $A_{E_1}$, corresponding to the $\mathbb{Z}_2$ symmetries associated with the elements: $-I$, $R_1$, $R_2$, $E_1$, $E_2$, respectively. The modification in the commutation relation would then be described by the following relation among the background fields: $\delta A_c = A_{R_1} \cup A_{E_1} + A_{R_2} \cup A_{E_2}$.   

\end{appendix}

\end{document}